\documentclass[journal]{IEEEtran}
\usepackage{cite}
\usepackage{amsmath,amssymb,amsfonts,amsthm}
\usepackage[utf8]{inputenc}

\usepackage{booktabs}
\usepackage{array}
\usepackage{xcolor}
\usepackage{graphicx}
\usepackage[caption=false,font=footnotesize]{subfig}
\usepackage{colortbl}
\usepackage{url}
\usepackage{float}
\usepackage{tikz}
\usetikzlibrary{arrows.meta,positioning,shapes.geometric}

\allowdisplaybreaks
\usepackage[linesnumbered,ruled,vlined]{algorithm2e}
\ifCLASSINFOpdf

\else

\fi

\begin{document}
\title{Site-specific Channel Modeling Based on Remote-Sensing Maps for 6G Space--Air--Ground Digital Twins}

\author{Peijie Liu, Pan Tang, Jianhua Zhang,~\IEEEmembership{Fellow,~IEEE}, Lei Tian, Bin Ao, Boyang He, and Hao Zheng% <-this % stops a space
	\thanks{This work was supported in part by National Natural Science Foundation of China (62341128, 62201086), National Key Research and Development Program of China (2023YFB2904805), Beijing Municipal Natural Fund (L243002) and Beijing University of Posts and Telecommunications-China Mobile Research Institute Joint Innovation Center.\textit{(Corresponding authors: Pan Tang; Jianhua Zhang.)}}% <-this % stops a space
	\thanks{Peijie Liu, Pan Tang, Jianhua Zhang, Lei Tian, Bin Ao, Boyang He and Hao Zheng are with the State
		Key Laboratory of Networking and Switching Technology, Beijing University of Posts and Telecommunications, Beijing 100876, China (e-mail: liupj@bupt.edu.cn; tangpan27@bupt.edu.cn; jhzhang@bupt.edu.cn; tianlbupt@bupt.edu.cn; binao@bupt.edu.cn; heboyang@bupt.edu.cn; zheng.hao@bupt.edu.cn).}% <-this % stops a space
	}

\markboth{}%
{Liu \MakeLowercase{\textit{et al.}}: Channel Modeling Based on Remote-Sensing Maps}

\maketitle

\begin{abstract}
Site-specific channel models are essential for wireless digital twins of sixth-generation (6G) space--air--ground communication systems. However, three-dimensional (3D) maps are difficult to obtain over wide areas, which limits large-area site-specific channel modeling. To address this issue, this paper proposes a remote-sensing-based augmented ray-tracing (RS-ART) channel modeling framework. The framework comprises a deterministic RT branch, a measurement-statistical branch, and an RT augmentation branch. To overcome the difficulty of acquiring large-area 3D maps, the deterministic RT branch reconstructs a 3D RT scene from satellite remote-sensing imagery and calibrates its electromagnetic material parameters using measured path loss. To provide the statistical parameters required for RT augmentation, the measurement-statistical branch establishes the marginal distributions and interparameter dependence models of the channel parameters. Specifically, a wideband unmanned aerial vehicle (UAV) channel measurement campaign is conducted at 4.60~GHz, and a proposed multipath estimation method estimates the complex amplitudes, delays, and Doppler shifts of the measured multipath. To bridge the gap between RT predictions and measurements, the RT augmentation branch organizes the RT multipath into line-of-sight (LoS), LoS-tail, and non-line-of-sight (NLoS) components, generates additional short-delay LoS-tail paths, and reallocates the component and path powers according to the measurement-derived statistics while preserving the total RT received power. The validation results show that the proposed framework reduces the path loss root-mean-square error (RMSE) from $5.45$ to $4.35$~dB and, relative to calibrated RT, decreases the RMS delay spread and normalized Doppler spread RMSEs by $53.03\%$ and $26.48\%$, respectively. The proposed framework provides a site-specific channel modeling approach for 6G space--air--ground digital-twin studies.
\end{abstract}


% \fi
\IEEEpeerreviewmaketitle

\section{Introduction}

Future sixth-generation (6G) communication systems are expected to extend wide-area coverage through satellite and unmanned aerial vehicle (UAV) communications~\cite{itu,wang2025intelligentsagin}. Satellite-to-ground channel research covers channel measurement, propagation characterization, modeling, and standardization over multiple frequency bands~\cite{suhong2026channel}. Compared with conventional terrestrial links, satellite and UAV links can exhibit longer propagation delays, larger Doppler shifts, and time-varying multipath determined by the local environment and platform motion~\cite{zhang2020channel,zhang2023channel,jiang2025ris}. Digital twin systems have been developed to emulate dynamic satellite networks and evaluate communication models~\cite{zhou2023satdt,gao2024plotinus}. At the channel level, wireless digital twins relate physical wireless environments to their corresponding channel responses~\cite{wang2025dtc,cai2026dtcsi,zhang2026dtmmwave}. Establishing this relation for a particular area requires site-specific channel models that account for the link locations and surrounding environment. Ray tracing (RT) is well suited to this task because it calculates propagation paths from three-dimensional (3D) geometry and electromagnetic material properties~\cite{hoydis2023sionna}.

RT has been used for site-specific channel analysis in satellite, UAV, and other outdoor communication scenarios. For satellite-to-ground links, the urban RT study in~\cite{cenni2024satrt} identifies nonspecular reflection as an important NLoS mechanism and reports that the Rician K-factor varies with satellite elevation. The suburban LEO study in~\cite{khawaja2026leort} further models elevation-dependent fading, terrain shadowing, and ground-antenna misalignment. For air-to-ground links, RT results from several urban environments show that LoS probability and path loss parameters vary with UAV altitude and building distribution~\cite{song2022a2grt}. The study in~\cite{liu2024maritimert} develops a measurement-corrected 3D RT model and analyzes the effects of environmental conditions, carrier frequency, and propagation distance on received power. These studies indicate that RT can represent environment- and position-dependent propagation effects under different link conditions.
In addition, measurement-assisted site-specific channel models have been developed by combining RT-resolved paths with measurement-derived or stochastic multipath. The hybrid model in~\cite{steinbock2016hybrid} combines deterministic RT components with a propagation graph to reproduce the diffuse tail observed in the measured delay and azimuth-delay power spectra. A related semi-deterministic model combines RT specular paths with graph-based diffuse scattering and obtains measurement-consistent delay and angular characteristics~\cite{tian2016semideterministic}. At low-terahertz frequencies, measured multipath clusters are matched to RT paths, while statistical components describe the remaining channel characteristics; the resulting model provides closer temporal and spatial statistics than conventional statistical and geometry-based stochastic models~\cite{chen2021hybrid}. Measurement-derived multipath classification has also been combined with RT geometry to separate target- and environment-related components for terahertz sensing and scene reconstruction~\cite{lyu2025hybrid}. These results show that retaining RT-resolved propagation paths while supplementing multipath not represented by the RT scene can improve the modeled channel statistics.

These RT applications require a 3D map of the propagation environment as input. OpenStreetMap (OSM) building data have been used to construct outdoor RT scenes and support automated propagation analysis~\cite{abouamer2025uncalibrated,ruznieto2023framework}. Although OSM provides accessible building footprints, many buildings lack height attributes and require additional height estimation~\cite{bernard2022osmheight}. Oblique aerial photography and high-precision 3D mapping can provide building and terrain geometry~\cite{li2022oap,zhang2024mapping}, whereas point-cloud measurements can support locally reconstructed indoor scenes~\cite{jarvelainen2016pointcloud,okamura2022pointcloud}. However, these data sources generally require existing geographic databases or dedicated local data acquisition, making complete 3D map inputs difficult to obtain over a wide area.

The limited availability of 3D maps restricts large-area site-specific RT modeling. In addition, a 3D RT scene may omit local and time-varying objects that contribute measurable multipath. To address these issues, this paper proposes the remote-sensing-based augmented ray-tracing (RS-ART) channel model, which constructs the RT input from remote-sensing imagery and uses measurement-derived statistics to supplement the RT results. The main contributions are summarized as follows.
\begin{itemize}
    \item The RS-ART channel modeling framework is proposed with deterministic RT, measurement-statistical, and RT augmentation branches. It provides a unified approach that constructs a 3D RT scene from satellite remote-sensing imagery, calibrates the electromagnetic material parameters, and incorporates measurement-derived channel statistics. This approach reduces the dependence on pre-existing 3D maps and simplifies RT scene acquisition.
    \item A multipath extraction method is proposed to jointly estimate the complex amplitudes, delays, and Doppler shifts of multipath components. It provides a joint parameter-estimation method for channel measurements in mobile scenarios, and the extracted paths support the subsequent statistical modeling.
    \item An RT augmentation method is proposed to reduce the differences between RT results and measured channel statistics. It provides a measurement-based approach to correcting RT multipath by generating additional short-delay LoS-tail paths and reallocating component and path powers according to the measured power ratios and delay and Doppler spreads.
    \item The RS-ART framework is validated using a $4.60$~GHz UAV channel measurement campaign and position-matched RT results. RS-ART reduces the path loss root-mean-square error (RMSE) from $5.45$ to $4.35$~dB and decreases the RMS delay spread and normalized Doppler spread RMSEs by $53.03\%$ and $26.48\%$, respectively.
\end{itemize}

The remainder of this paper is organized as follows. Section II introduces the three branches of the RS-ART framework and the overall channel composition. Section III presents the channel measurement and proposed multipath extraction method. Section IV describes the deterministic RT branch, including scene construction, RT path classification, and electromagnetic parameter calibration. Section V presents the measurement-derived statistical models and fitting results. Section VI describes the RT augmentation branch and validates the resulting channels. Section VII concludes the paper.

\section{RS-ART Three-Branch Framework}
\label{sec:framework}

\begin{figure*}[ht]
    \centering
    \includegraphics[width=0.9\textwidth]{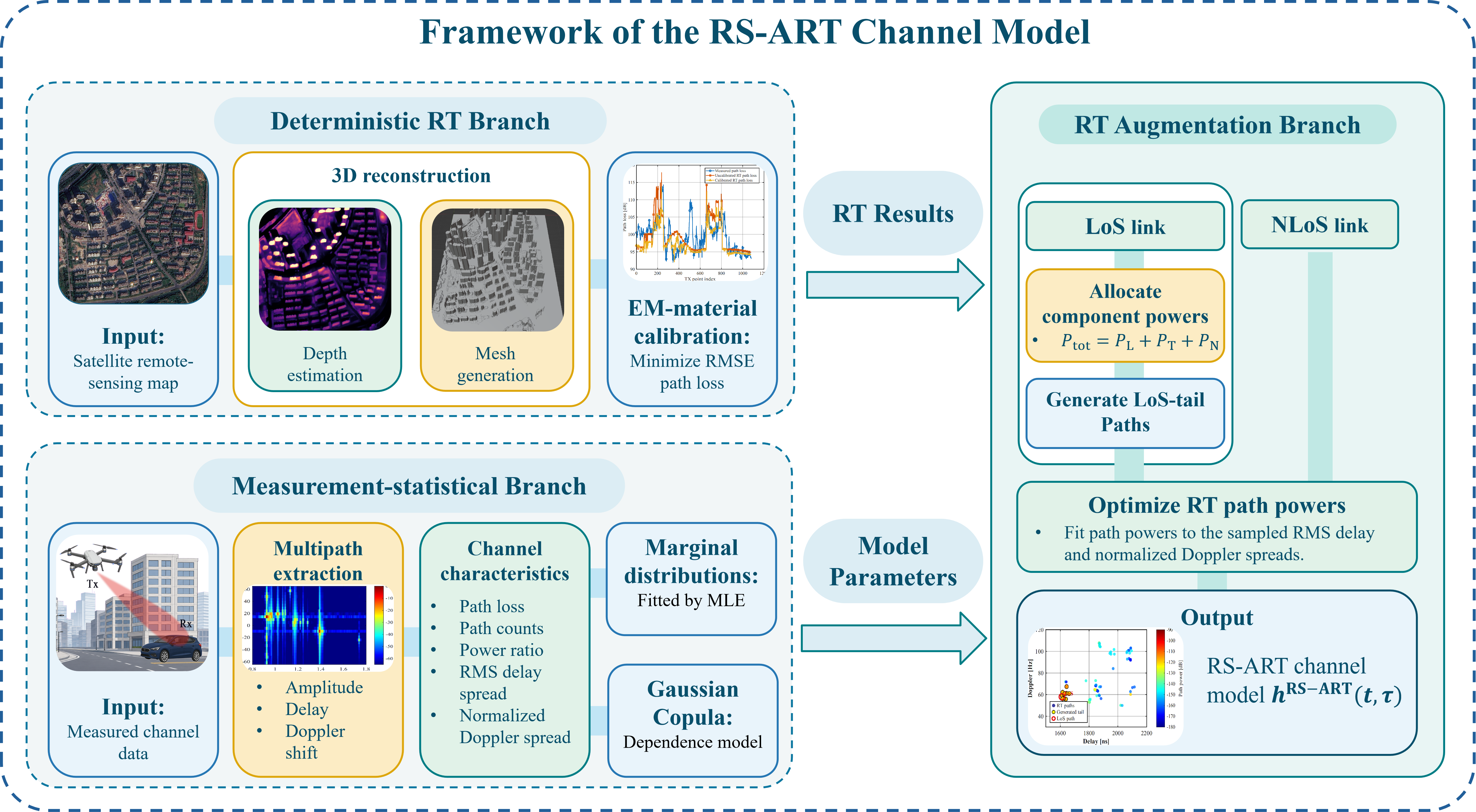}
    \caption{Information flow of the three-branch RS-ART channel modeling framework.}
    \label{fig:rs_art_framework}
\end{figure*}

This section presents the information flow of the RS-ART framework. As shown in Fig.~\ref{fig:rs_art_framework}, the framework consists of a deterministic RT branch, a measurement-statistical branch, and an RT augmentation branch. These branches address, in sequence, the acquisition of a 3D RT input, the estimation of measurement-derived model parameters, and the use of these parameters to augment the RT channel.

\subsection{Deterministic RT Branch}

Acquiring complete 3D maps over a wide area through dedicated surveying is costly. The deterministic RT branch therefore uses readily available satellite remote-sensing imagery as its scene input. Depth estimation and mesh generation convert the imagery into a 3D RT scene, after which the electromagnetic material parameters are calibrated using measured path loss. This branch outputs the RT path coefficients, delays, Doppler shifts, LoS/NLoS state, and total received power for each transmitter (Tx)--receiver (Rx) link. The scene construction, RT path classification, and calibration are presented in Section~\ref{sec:rt_construction}.

\subsection{Measurement-Statistical Branch}

The measurement-statistical branch obtains channel samples for establishing the statistical models required by RT augmentation. The proposed multipath extraction method estimates the complex amplitudes, delays, and Doppler shifts of the measured paths, from which the path counts, component power ratios, RMS delay spreads, and normalized Doppler spreads are calculated. The marginal distribution of each parameter is fitted by maximum likelihood estimation, and Gaussian copula models describe the dependence among parameters within each group. The extracted multipath coefficients are also used to calculate the path loss for calibrating the electromagnetic material parameters in the deterministic RT branch. The resulting marginal distributions and dependence models provide the statistical inputs to the RT augmentation branch. The measurement and parameter extraction are presented in Section~\ref{sec:measurement_extraction}, and the fitted statistical models are presented in Section~\ref{sec:statistical_modeling}.

\subsection{RT Augmentation Branch and Channel Composition}
\label{sec:channel_model}

The RT augmentation branch combines the RT results with parameters sampled from the measurement-derived models. A comparison between the measured and RT channels indicates that RT does not fully represent the short-delay multipath near the LoS path. Such multipath is mainly associated with nearby and time-varying scatterers, including pedestrians, vegetation, and vehicles, that are not explicitly represented in the 3D RT scene. RS-ART therefore organizes the channel into LoS, LoS-tail, and NLoS components according to the LoS/NLoS propagation condition:
\begin{equation}
	h^{\mathrm{RS\text{-}ART}}(t,\tau)
	=
	I_{\mathrm{LoS}} h_{\mathrm L}(t,\tau)
	+
	I_{\mathrm{LoS}} h_{\mathrm T}(t,\tau)
	+
	h_{\mathrm N}(t,\tau),
	\label{eq:rs_art_channel}
\end{equation}
where $I_{\mathrm{LoS}}=1$ for LoS links and $I_{\mathrm{LoS}}=0$ for NLoS links. Here, $h_{\mathrm L}(t,\tau)$ denotes the deterministic LoS component, $h_{\mathrm T}(t,\tau)$ denotes the LoS-tail component, and $h_{\mathrm N}(t,\tau)$ denotes the residual-NLoS component for a LoS link and the full NLoS channel for an NLoS link. Thus, LoS links include all three components, whereas NLoS links retain only $h_{\mathrm N}(t,\tau)$.

Each component follows the common path-sum representation
\begin{equation}
    h_c(t,\tau)
    =
    \sum_{p\in\mathcal S_c}
    a_p^{c}
    e^{j2\pi\nu_p^{c}t}
    \delta\left(\tau-\tau_p^{c}\right),
\end{equation}
where $c\in\{\mathrm L,\mathrm T,\mathrm N\}$, $\mathcal S_c$ is the corresponding path set, and $a_p^{c}$, $\tau_p^{c}$, and $\nu_p^{c}$ are the complex coefficient at $t=0$, delay, and Doppler shift of path $p$, respectively. Here, $\delta(\cdot)$ is the Dirac delta function.

\begin{figure}[H]
    \centering
    \includegraphics[width=0.95\columnwidth]{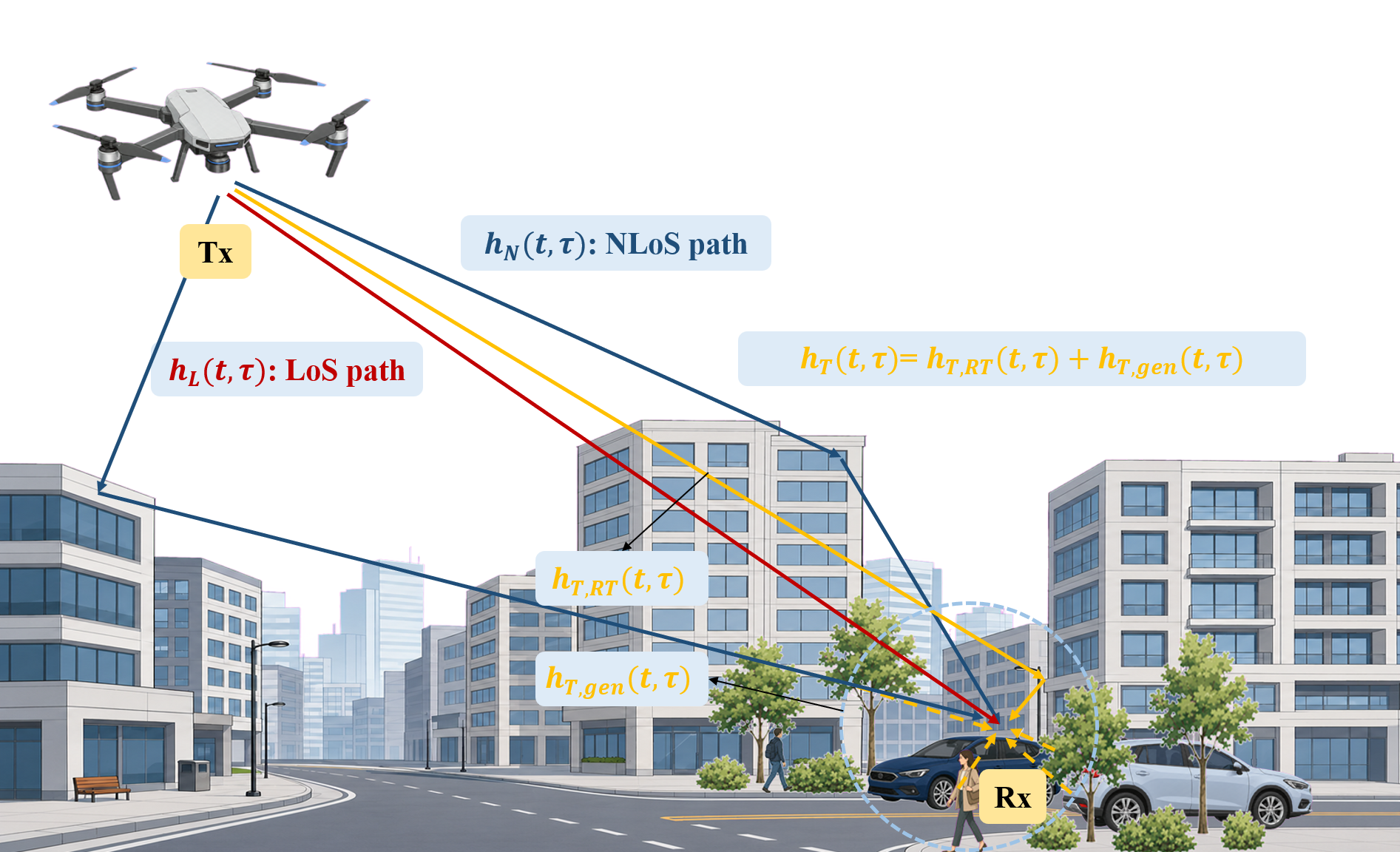}
    \caption{Physical origins of the RS-ART channel components in an air-to-ground scenario.}
    \label{fig:rs_art_component_structure}
\end{figure}

Fig.~\ref{fig:rs_art_component_structure} illustrates the physical origins of these components. The direct path forms $h_{\mathrm L}(t,\tau)$ and provides the LoS reference delay and Doppler. The LoS-tail component contains an RT-resolved part and a statistically generated part:
\begin{equation}
    h_{\mathrm T}(t,\tau)
    =
    h_{\mathrm{T,RT}}(t,\tau)
    +
    h_{\mathrm{T,gen}}(t,\tau).
\end{equation}
Here, $h_{\mathrm{T,RT}}(t,\tau)$ contains the short-delay paths resolved by RT from the reconstructed buildings, ground, and other modeled objects. The statistically generated part $h_{\mathrm{T,gen}}(t,\tau)$ represents additional short-delay multipath associated with local objects that are difficult to include in the RT scene, such as vehicles, roadside facilities, and vegetation near the Rx. Their positions and electromagnetic properties can vary during measurement, and their proximity to the Rx can result in measurable received power.

For an RT-LoS link, the remaining RT paths outside the LoS-tail delay region form $h_{\mathrm N}(t,\tau)$. For an RT-NLoS link, no LoS or LoS-tail component is constructed, and $h_{\mathrm N}(t,\tau)$ represents the full channel. Both LoS-tail parts are defined relative to the RT LoS reference and restricted to a finite excess-delay region whose upper bound is denoted by $\tau_{\mathrm T}$. The augmentation branch generates additional LoS-tail paths when required and reallocates the component and path powers according to the sampled power ratios and delay and Doppler spreads while preserving the total RT received power. The RT path classification is presented in Section~\ref{sec:rt_construction}, while the augmentation models, realization procedure, and validation are presented in Section~\ref{sec:validation}.

\section{Channel Measurement and Parameter Extraction Method}
\label{sec:measurement_extraction}

This section presents the UAV measurement campaign, the correlation-based construction of the measured channel response, and the proposed method for extracting the complex amplitudes, delays, and Doppler shifts of the multipath components.
\subsection{Measurement System and Campaign}

\begin{figure}[htb]
    \centering
    \includegraphics[width=\columnwidth]{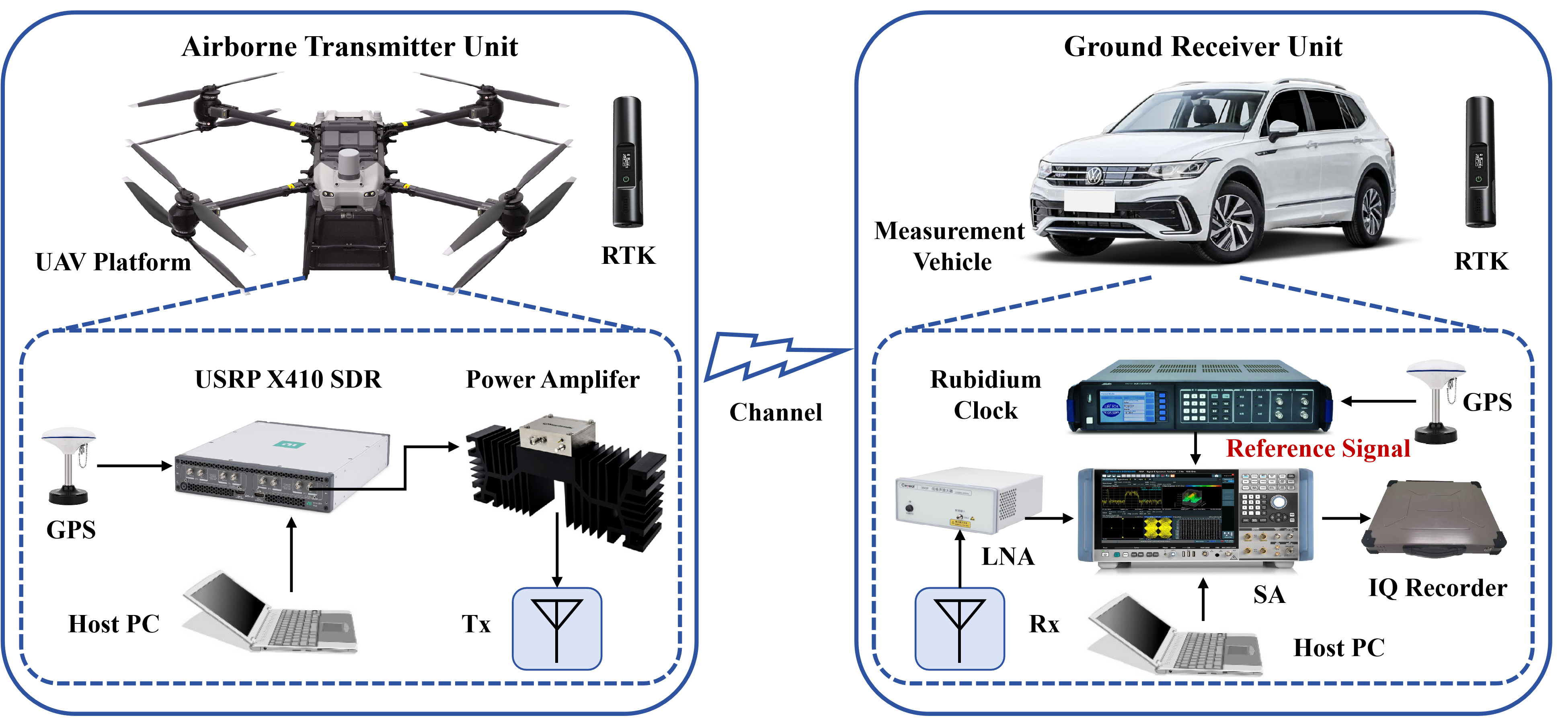}
    \caption{Channel measurement platform.}
    \label{fig:measurement_platform}
\end{figure}

The channel sounding platform is shown in Fig.~\ref{fig:measurement_platform}. The airborne Tx unit is mounted on a UAV platform. A host computer controls the Universal Software Radio Peripheral (USRP) X410 software-defined radio to generate the binary phase-shift keying (BPSK)-modulated pseudo-noise (PN) sounding waveform, and the generated radio-frequency signal is amplified by the power amplifier before being radiated by the Tx antenna. A Global Positioning System (GPS) module provides the frequency reference required by the Tx chain, while a real-time kinematic (RTK) module records the UAV position and altitude at a 10~Hz update rate. The GPS and RTK records provide the Tx frequency reference and trajectory used in post-processing. The recorded samples are stored for subsequent channel impulse response (CIR) estimation.

The ground Rx unit is deployed on the measurement vehicle and remains fixed during the mobile measurement. The received signal is captured by the Rx antenna, amplified by the low-noise amplifier, and then downconverted and sampled by the spectrum analyzer (SA). The in-phase/quadrature (IQ) recorder stores the sampled complex IQ data for offline correlation and multipath parameter extraction. As shown in Fig.~\ref{fig:measurement_platform}, the Rx-side reference is provided by a GPS-disciplined rubidium clock, whose reference signal is fed to the SA and Rx-side recording chain. An RTK module records the ground Rx position. The rubidium-clock reference is used for received IQ sampling and subsequent Doppler estimation. Table~\ref{tab:measurement_configurations} summarizes the measurement configuration. The sounder operates at $4.60$~GHz with a $250$~MHz bandwidth and sampling rate, and uses a BPSK-modulated PN sequence of length $N_{\mathrm c}=511$. The RTK update rate is $10$~Hz, the UAV altitude is approximately $150$~m, and the Rx antenna height is approximately $1.80$~m.

\begin{table}[htb]
    \caption{Measurement system configurations.}
    \label{tab:measurement_configurations}
    \centering
    \renewcommand{\arraystretch}{1.15}
    \begin{tabular}{cc}
        \toprule
        Configuration & Value \\
        \midrule
        Center frequency & 4.60 GHz \\
        Bandwidth & 250 MHz \\
        Sampling rate & 250 MHz \\
        Modulation scheme & BPSK \\
        Sounding signal & PN sequence \\
        PN sequence length & $N_{\mathrm c}=511$ \\
        RTK update rate & 10 Hz \\
        UAV flight altitude & $\sim$150 m \\
        Rx antenna height & $\sim$1.80 m \\
        \bottomrule
    \end{tabular}
\end{table}

\begin{figure}[htb]
    \centering
    \begin{tabular}{@{}c@{\hspace{0.004\columnwidth}}c@{}}
        \includegraphics[height=0.498\columnwidth]{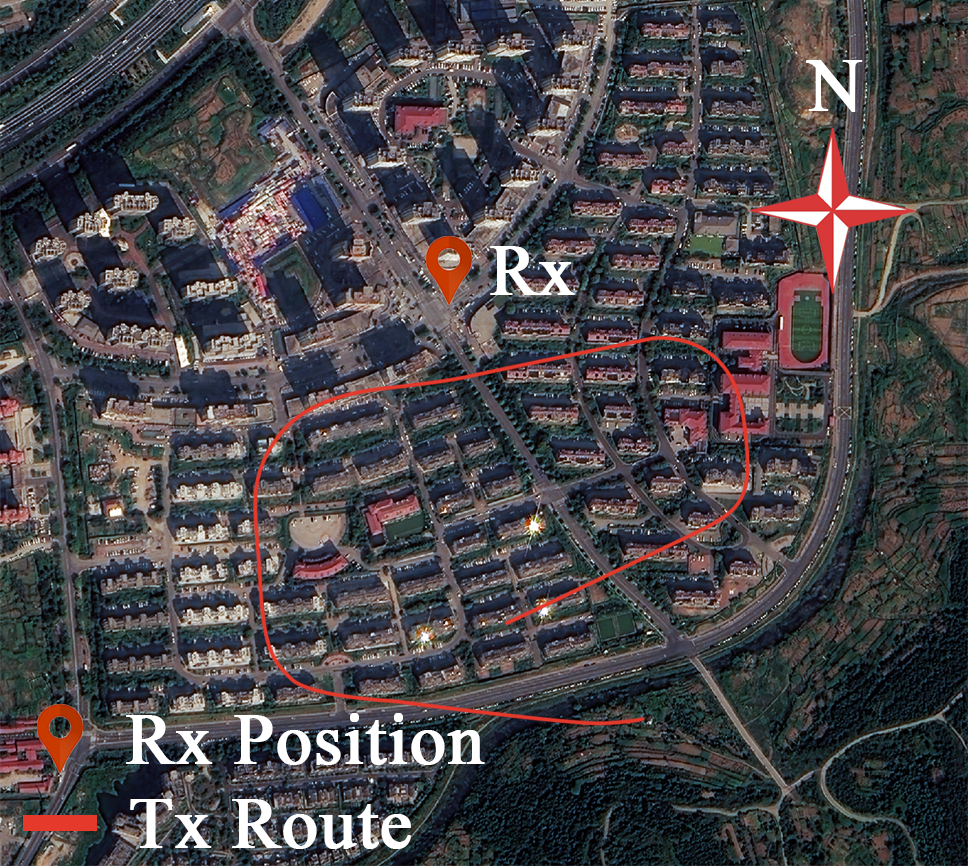} &
        \includegraphics[height=0.498\columnwidth]{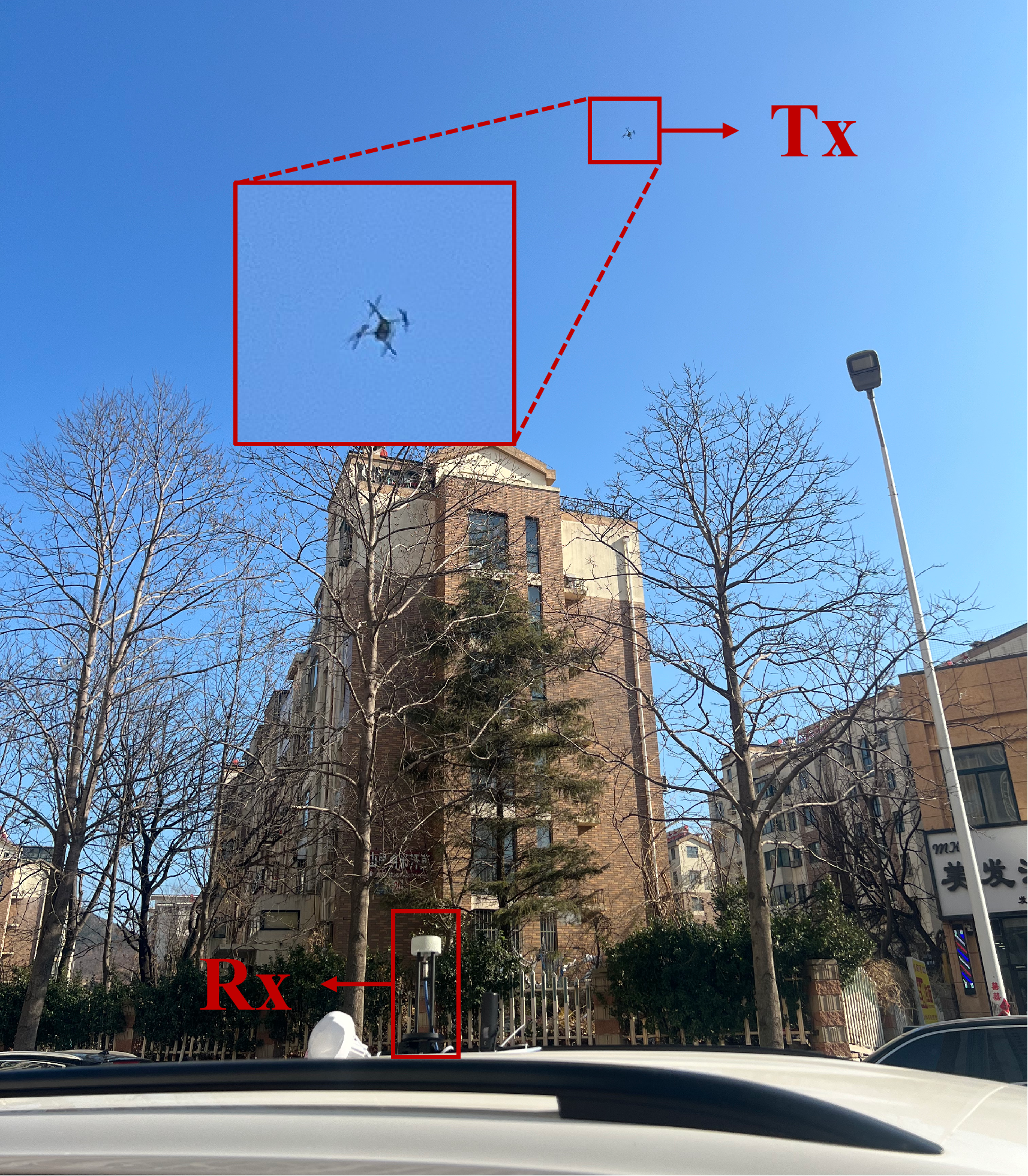} \\
        (a) & (b)
    \end{tabular}
    \caption{Measurement campaign: (a) route and (b) site photograph.}
    \label{fig:measurement_campaign}
\end{figure}

The campaign adopts a mobile-measurement scheme with a fixed ground Rx and a moving airborne Tx. Fig.~\ref{fig:measurement_campaign}(a) shows the measurement route, where the Rx position is fixed on the ground and the UAV-borne Tx moves along the marked trajectory at an altitude of approximately 150 m. The airborne RTK records the Tx position and altitude at 10~Hz. Accordingly, measurement frames are extracted at the same 0.10~s interval and associated with the RTK positions through their timestamps, while the Rx-side RTK record provides the fixed Rx location. Fig.~\ref{fig:measurement_campaign}(b) gives the corresponding ground-level measurement photo, in which the airborne Tx and the ground Rx are indicated.

During data acquisition, the IQ recorder continuously samples the received signal for approximately two minutes at $f_{\mathrm s}=250$~MHz. The fast-time interval is therefore $T_{\mathrm f}=1/f_{\mathrm s}=4.00$~ns. Each CIR contains $N_{\mathrm f}=1022$ samples and has duration $T_{\mathrm{CIR}}=N_{\mathrm f}T_{\mathrm f}=4.088~\mu\mathrm{s}$. One CIR is retained every 50 CIR periods, giving the slow-time interval $T_{\mathrm s}=50T_{\mathrm{CIR}}=204.40~\mu\mathrm{s}$. Every $N_{\mathrm s}=1024$ retained CIRs form one frame of duration $T_{\mathrm{frame}}=N_{\mathrm s}T_{\mathrm s}=0.2093$~s. Successive frames are shifted by 0.10~s to follow the 10~Hz RTK position updates and therefore overlap in time. The two-minute recording provides approximately 1200 candidate frames before timestamp matching and validity screening.

\subsection{Measurement Principle}

Each CIR is obtained through time-domain correlation between a received IQ snapshot and the direct-connected PN calibration reference~\cite{tang2021thzmeasurement}. In the implementation below, this correlation is evaluated equivalently by frequency-domain multiplication followed by an inverse discrete Fourier transform (DFT). A DFT along slow time then constructs the calibration-referenced two-dimensional delay--Doppler (DD) channel response.

The wideband sounder transmits a periodically repeated BPSK PN probing waveform. Let
$b[q]\in\{\pm1\}$ denote one period of the PN sequence, where
$q=0,\ldots,N_{\mathrm c}-1$, and let $T_{\mathrm c}$ be the chip interval. The equivalent complex baseband probing waveform over one period is
\begin{equation}
	x(t)
	=
	\sum_{q=0}^{N_{\mathrm c}-1}
	b[q]p_{\mathrm c}(t-qT_{\mathrm c}),
	\label{eq:tx_pn_waveform}
\end{equation}
where $p_{\mathrm c}(t)$ is the chip pulse.

After downconversion, the received complex baseband signal is modeled as
\begin{equation}
	y(t)
	=
	\int h(t,\tau)x(t-\tau)\,\mathrm d\tau
	+
	n_{\mathrm{Rx}}(t),
	\label{eq:rx_signal_model}
\end{equation}
where $h(t,\tau)$ is the time-varying channel impulse response on the physical delay axis $\tau$, and $n_{\mathrm{Rx}}(t)$ is the Rx noise. Channel variation over the fast-time samples within each retained snapshot is assumed to be negligible, whereas channel evolution across retained snapshots is preserved along slow time. This quasi-static assumption within one PN period allows the fast-time samples to be processed as one delay observation. For one measurement frame, the received samples are therefore arranged into
$\mathbf S_{\mathrm{mea}}\in\mathbb C^{N_{\mathrm f}\times N_{\mathrm s}}$ as
\begin{equation}
	[\mathbf S_{\mathrm{mea}}]_{n,m}
	=
	y(mT_{\mathrm s}+nT_{\mathrm f}),
	\label{eq:received_sample_matrix}
\end{equation}
where $n=0,\ldots,N_{\mathrm f}-1$ and $m=0,\ldots,N_{\mathrm s}-1$.

Prior to the propagation measurement, the Tx and Rx ports are directly connected to record the calibration signal $y_{\mathrm{cal}}(t)$. Let $s_{\mathrm{cal}}[n]=y_{\mathrm{cal}}(nT_{\mathrm f})$ denote its fast-time samples. The same calibration reference is applied to every retained snapshot; hence, the sampled vector is replicated along the slow-time dimension as
\begin{equation}
	\begin{aligned}
	\mathbf s_{\mathrm{cal}}
	&=
	[s_{\mathrm{cal}}[0],s_{\mathrm{cal}}[1],\ldots,
	s_{\mathrm{cal}}[N_{\mathrm f}-1]]^{\mathrm T},\\
	\mathbf S_{\mathrm{cal}}
	&=
	\mathbf s_{\mathrm{cal}}\mathbf 1_{N_{\mathrm s}}^{\mathrm T},
	\end{aligned}
	\label{eq:calibration_matrix}
\end{equation}
where $\mathbf 1_{N_{\mathrm s}}$ is the length-$N_{\mathrm s}$ all-ones vector. Let $\mathcal F_N\{\cdot\}$ and $\mathcal F_N^{-1}\{\cdot\}$ denote the $N$-point DFT and inverse DFT, respectively, and let $\mathbf F_N$ be the corresponding unnormalized DFT matrix, with $[\mathbf F_N]_{u,v}=\exp(-j2\pi uv/N)$. Correlating each received snapshot with the calibration reference in the fast-time frequency domain gives the delay--time observation
\begin{equation}
	\mathbf Z
	=
	\frac{
		\mathbf F_{N_{\mathrm f}}^{\mathrm H}
		\left[
		\left(\mathbf F_{N_{\mathrm f}}\mathbf S_{\mathrm{mea}}\right)
		\odot
		\left(\mathbf F_{N_{\mathrm f}}\mathbf S_{\mathrm{cal}}\right)^{*}
		\right]
	}{
		N_{\mathrm f}\left\|\mathbf s_{\mathrm{cal}}\right\|_{2}^{2}
	},
	\label{eq:matrix_delay_time_measurement}
\end{equation}
where $\mathbf Z\in\mathbb C^{N_{\mathrm f}\times N_{\mathrm s}}$, $\odot$ denotes the Hadamard product, and $(\cdot)^*$ denotes complex conjugation. Thus, each column of $\mathbf Z$ is a correlation-derived delay profile for one retained snapshot.

To express the Doppler samples directly in centered order, define the slow-time DFT matrix $\mathbf F_{N_{\mathrm s}}^{\mathrm c}$ by $[\mathbf F_{N_{\mathrm s}}^{\mathrm c}]_{k,m}=\exp[-j2\pi(k-k_0)m/N_{\mathrm s}]$. Applying this transform along the slow-time dimension gives
\begin{equation}
\begin{aligned}
	\mathbf Y_{\tau\nu}
	&=
	\mathbf Z\left(\mathbf F_{N_{\mathrm s}}^{\mathrm c}\right)^{\mathrm T},\\
	\nu_k
	&=
	\frac{k-k_0}{N_{\mathrm s}T_{\mathrm s}},
	\quad k=0,\ldots,N_{\mathrm s}-1,
\end{aligned}
	\label{eq:matrix_dd_measurement}
\end{equation}
where $k_0=N_{\mathrm s}/2$ for the even value of $N_{\mathrm s}$ used here.

The correlation output retains the finite delay resolution of the sounding waveform. To include this response in the subsequent multipath dictionary, define the normalized circular autocorrelation of the sampled calibration signal as
\begin{equation}
	r_{\mathrm{cal}}[n]
	=
	\frac{
		N_{\mathrm f}
		\mathcal F_{N_{\mathrm f}}^{-1}
		\left\{
		\left|\widetilde{s}_{\mathrm{cal}}[\mu]\right|^2
		\right\}[n]
	}{
		\sum_{\mu=0}^{N_{\mathrm f}-1}
		\left|\widetilde{s}_{\mathrm{cal}}[\mu]\right|^2
	},
	\label{eq:discrete_calibration_pulse}
\end{equation}
where $\widetilde{s}_{\mathrm{cal}}[\mu]=\mathcal F_{N_{\mathrm f}}\{s_{\mathrm{cal}}[n]\}[\mu]$ and $\mu$ is the fast-time frequency-bin index. Let $\mathbf R_{\mathrm{cal}}\in\mathbb C^{N_{\mathrm f}\times N_{\mathrm f}}$ be the corresponding circulant matrix, with $[\mathbf R_{\mathrm{cal}}]_{n,n'}=r_{\mathrm{cal}}[(n-n')_{N_{\mathrm f}}]$, where $(\cdot)_N$ denotes modulo-$N$ indexing. If $\mathbf H_{\mathrm{DT}}$ denotes the sampled delay--time channel and $\mathbf H_{\tau\nu}$ denotes its slow-time DFT, the measured responses satisfy
\begin{equation}
	\begin{aligned}
	\mathbf Z
	&\approx
	\mathbf R_{\mathrm{cal}}\mathbf H_{\mathrm{DT}}
	+
	\mathbf E,\\
	\mathbf H_{\tau\nu}
	&=
	\mathbf H_{\mathrm{DT}}
	\left(\mathbf F_{N_{\mathrm s}}^{\mathrm c}\right)^{\mathrm T},\\
	\mathbf Y_{\tau\nu}
	&\approx
	\mathbf R_{\mathrm{cal}}\mathbf H_{\tau\nu}
	+
	\mathbf W_{\tau\nu},
	\end{aligned}
	\label{eq:measured_true_dd_relation}
\end{equation}
where $\mathbf E$ is the post-correlation noise matrix and $\mathbf W_{\tau\nu}=\mathbf E(\mathbf F_{N_{\mathrm s}}^{\mathrm c})^{\mathrm T}$. Equation~\eqref{eq:measured_true_dd_relation} shows that the measured response is convolved with $r_{\mathrm{cal}}[n]$ along delay and corrupted by measurement noise. Thus, $\mathbf Y_{\tau\nu}$ provides the measured delay--Doppler response, while $r_{\mathrm{cal}}[n]$ provides the calibration pulse used in the multipath dictionary.

A path whose Doppler shift falls between two Doppler-grid points spreads over multiple bins after the finite-length slow-time DFT. This broadening results from the finite observation window and the mismatch between the physical Doppler shift and the discrete grid. The measured DD response therefore includes delay-domain broadening from $r_{\mathrm{cal}}[n]$ and Doppler-domain broadening from finite-window spectral leakage.

\subsection{Proposed Multipath Extraction Method}

The correlation-derived channel response is converted into continuous multipath parameters in two stages. A two-dimensional cell-averaging constant false-alarm rate (CA-CFAR) detector~\cite{rohling1983cfar}, a local-maximum test, and a Doppler-support mask first select candidate bins from the measured power map. Orthogonal matching pursuit (OMP)~\cite{tropp2007omp} then refines their delays and Doppler shifts on local grids and estimates their complex coefficients.

The coarse-estimation stage identifies approximate delay and Doppler locations for subsequent sparse refinement. The CA-CFAR detector is applied to the measured power map
\begin{equation}
    U[n,k]=\left|[\mathbf Y_{\tau\nu}]_{n,k}\right|^{2},
\end{equation}
where $n$ is the delay-bin index corresponding to the nominal delay $\tau_n=nT_{\mathrm f}$, and $k$ is the centered Doppler-bin index corresponding to $\nu_k$. For a cell under test $(n,k)$, let $\mathcal{T}_{n,k}$ be the set of training cells around the cell after excluding the guard region. The local background power and CA-CFAR threshold are
\begin{equation}
\begin{aligned}
    \widehat U_{\mathrm{bg}}[n,k]
    &=
    \frac{1}{|\mathcal{T}_{n,k}|}
    \sum_{(a,b)\in\mathcal{T}_{n,k}}
    U[a,b],\\
    \Gamma_{n,k}
    &=
    |\mathcal{T}_{n,k}|
    \left(
    P_{\mathrm{FA}}^{-1/|\mathcal{T}_{n,k}|}-1
    \right)
    \widehat U_{\mathrm{bg}}[n,k],
\end{aligned}
\end{equation}
where $P_{\mathrm{FA}}$ is the target false-alarm probability. In the implementation, the guard region contains four delay bins and three Doppler bins on each side of the cell under test, while the training region contains 16 delay bins and 10 Doppler bins on each side outside the guard region. The false-alarm probability is set to $P_{\mathrm{FA}}=10^{-5}$. A bin is retained only if $U[n,k]>\Gamma_{n,k}$ and it is a local maximum in a $5\times 5$ neighborhood. To remove detections outside the observed Doppler range, an additional support mask is centered on zero Doppler and its half-width is determined from the strongest detected Doppler offset:
\begin{equation}
    \left|\nu_k\right|
    \leq
    \max\left(
    10\left|\nu_{k_{\max}}\right|,
    3\Delta\nu
    \right),
    \quad
    \Delta\nu=\frac{1}{N_{\mathrm s}T_{\mathrm s}},
    \label{eq:doppler_support_mask}
\end{equation}
where $k_{\max}$ is the Doppler index of the strongest detected bin. Sorting the retained bins in descending order of $U[n,k]$ gives the coarse candidate set $\mathcal{C}_0$ in the physical coordinates $(\tau_n,\nu_k)$.

The coarse bins initialize continuous parameter refinement. Each measured path is represented by $(a_p^{\mathrm{meas}},\tau_p^{\mathrm{meas}},\nu_p^{\mathrm{meas}})$, where the three entries denote its complex coefficient, propagation delay, and Doppler shift, respectively. This notation distinguishes the measured path parameters from the RT parameters $(a_i^{\mathrm{RT}},\tau_i^{\mathrm{RT}},\nu_i^{\mathrm{RT}})$ in Section~\ref{sec:channel_model}. For a candidate pair $(\tau,\nu)$, a continuous delay shift is applied to the measured calibration response and the corresponding delay--slow-time atom is
\begin{equation}
\begin{aligned}
    \varphi[n;\tau]
    &=
    \mathcal{F}_{N_{\mathrm{f}}}^{-1}
    \left\{
    \widetilde r_{\mathrm{cal}}[\mu]
    \exp\left(-j2\pi f_{\mu}\tau\right)
    \right\}[n],\\
    B[n,m;\tau,\nu]
    &=
    \varphi[n;\tau]
    \exp\left(j2\pi \nu mT_{\mathrm{s}}\right),
\end{aligned}
\end{equation}
where $\widetilde r_{\mathrm{cal}}[\mu]=\mathcal{F}_{N_{\mathrm{f}}}\{r_{\mathrm{cal}}[n]\}[\mu]$, and $f_{\mu}$ is the standard fast Fourier transform (FFT)-ordered fast-time frequency grid with bin spacing $1/(N_{\mathrm f}T_{\mathrm f})$ and $\mu=0,\ldots,N_{\mathrm f}-1$. Here, $n$, $m$, and $\mu$ index delay samples, slow-time samples, and fast-time frequency bins, respectively.

After applying the same slow-time DFT and Doppler centering used to form $\mathbf Y_{\tau\nu}$, the resulting atom is vectorized as $\mathbf{d}(\tau,\nu)$. This transformed atom includes the delay response of the calibration pulse and the Doppler leakage pattern associated with an off-grid Doppler shift. With $\mathbf y_{\tau\nu}=\operatorname{vec}(\mathbf Y_{\tau\nu})$, the measured response is approximated by a sparse superposition of these atoms:
\begin{equation}
    \mathbf{y}_{\tau\nu}
    \approx
    \sum_{p=1}^{N_{\mathrm{est}}}
    a_p^{\mathrm{meas}}
    \mathbf{d}(\tau_p^{\mathrm{meas}},\nu_p^{\mathrm{meas}}),
\end{equation}
where $N_{\mathrm{est}}$ is the number of estimated paths. Using $\mathcal{C}_0$ as the initial candidate set, OMP refines each candidate over a local two-dimensional grid. At each iteration, the atom that produces the largest residual reduction is added to the selected support $\mathcal{S}$, and all selected complex coefficients are re-estimated by regularized least squares. With $\mathbf{D}_{\mathcal{S}}$ denoting the selected atom matrix, the coefficient update is
\begin{equation}
    \widehat{\mathbf{a}}_{\mathcal{S}}^{\mathrm{meas}}
    =
    \left(
    \mathbf{D}_{\mathcal{S}}^{\mathrm H}\mathbf{D}_{\mathcal{S}}
    +
    \lambda_{\mathrm{reg}}\mathbf{I}_{|\mathcal S|}
    \right)^{-1}
    \mathbf{D}_{\mathcal{S}}^{\mathrm H}\mathbf{y}_{\tau\nu},
\end{equation}
where $\lambda_{\mathrm{reg}}>0$ is the regularization coefficient and $\mathbf I_{|\mathcal S|}$ is the identity matrix of order $|\mathcal S|$. The residual is updated after each selection. For the measured data, OMP selection is capped at $N_{\max}=60$ candidates and terminates earlier if no candidate remains. After the coefficient update, only components satisfying
$10\log_{10}(|\widehat a_p^{\mathrm{meas}}|^2/\max_{q\in\mathcal S}|\widehat a_q^{\mathrm{meas}}|^2)\geq-40.00$~dB are retained. Therefore, $N_{\mathrm{est}}\leq N_{\max}$ denotes the actual number of retained paths in the frame, including the case in which fewer than 60 candidates are available. The resulting measured multipath set is
\begin{equation}
    \widehat{\mathcal{G}}_{\mathrm{meas}}
    =
    \left\{
    \left(
    \widehat a_p^{\mathrm{meas}},
    \widehat \tau_p^{\mathrm{meas}},
    \widehat \nu_p^{\mathrm{meas}}
    \right)
    \right\}_{p=1}^{N_{\mathrm{est}}}.
\end{equation}

Before processing the measurements, the extraction procedure is first checked with a synthetic response generated using the same calibration waveform and delay--slow-time sampling configuration as the measurement. A sparse channel with 15 off-grid multipath components is generated within $\tau\leq1.5~\mu\mathrm{s}$ and $|\nu|\leq300$~Hz. The path powers follow an exponential delay decay with an approximately 40~dB span between the strongest and weakest components, and additive complex Gaussian noise is included with a signal-to-noise ratio (SNR) of 15~dB. The maximum candidate count is set to 17, two more than the injected path count, to include additional estimates caused by off-grid mismatch or noise.

\begin{figure}[htbp]
    \centering
    \subfloat[]{\includegraphics[width=0.5\columnwidth]{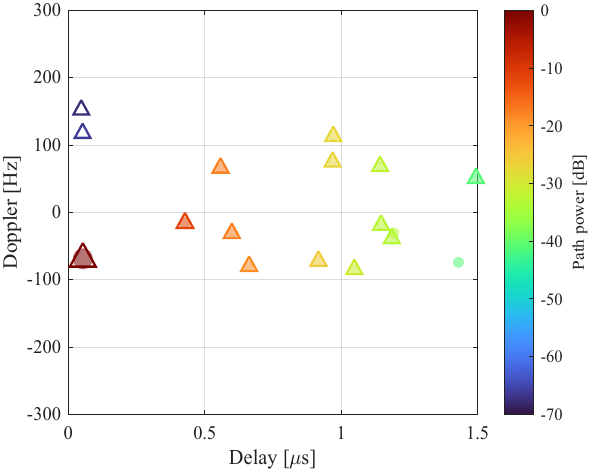}}
    \hfill
    \subfloat[]{\includegraphics[width=0.5\columnwidth]{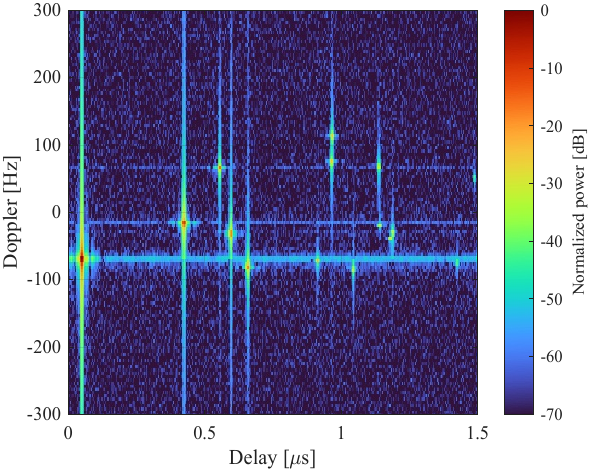}}\\[0.6em]
    \subfloat[]{\includegraphics[width=0.5\columnwidth]{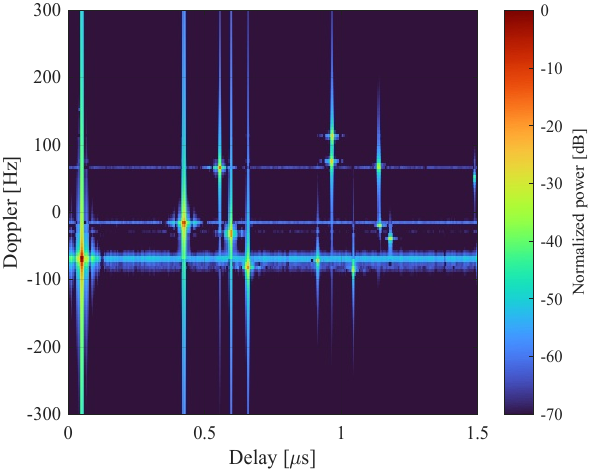}}
    \hfill
    \subfloat[]{\includegraphics[width=0.5\columnwidth]{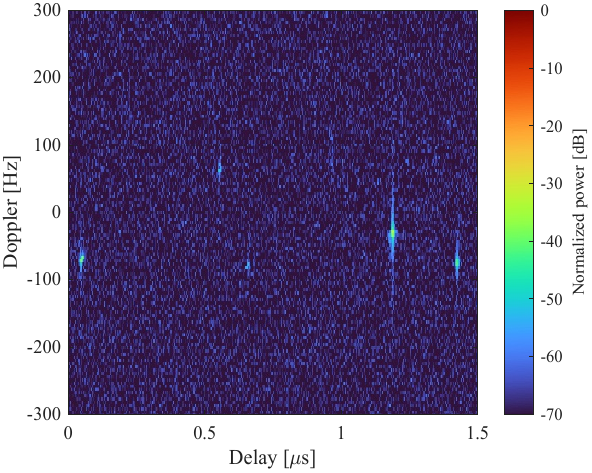}}
    \caption{Synthetic validation of the sparse parameter extraction method: (a) true and estimated paths; (b) noisy response; (c) reconstruction; and (d) residual.}
    \label{fig:synthetic_dd_extraction}
\end{figure}

Fig.~\ref{fig:synthetic_dd_extraction}(a) compares the true and estimated path parameters, while Fig.~\ref{fig:synthetic_dd_extraction}(b)--(d) show the noisy calibration-referenced response, sparse reconstruction, and residual, respectively. All power maps are normalized by the maximum response power and expressed in decibels. The estimated markers overlap the injected locations for the high-power paths. Two low-power estimates occur beyond the imposed $\tau\leq1.5~\mu\mathrm{s}$ support, while one weak component near $(1.43~\mu\mathrm{s},-75~\mathrm{Hz})$ is not matched. At an SNR of $15$~dB, the procedure identifies most high-power components, while errors remain for weak and off-grid paths.

\begin{figure*}[htbp]
    \centering
    \subfloat[]{\includegraphics[width=0.64\columnwidth]{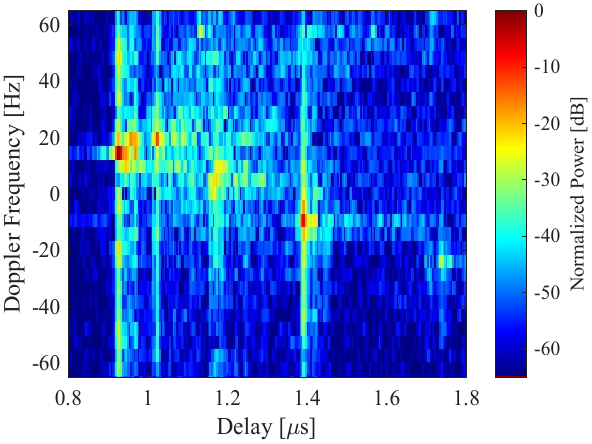}}
    \hfill
    \subfloat[]{\includegraphics[width=0.64\columnwidth]{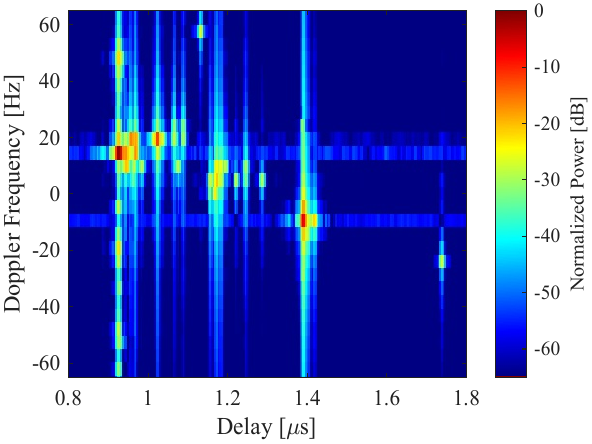}}
    \hfill
    \subfloat[]{\includegraphics[width=0.64\columnwidth]{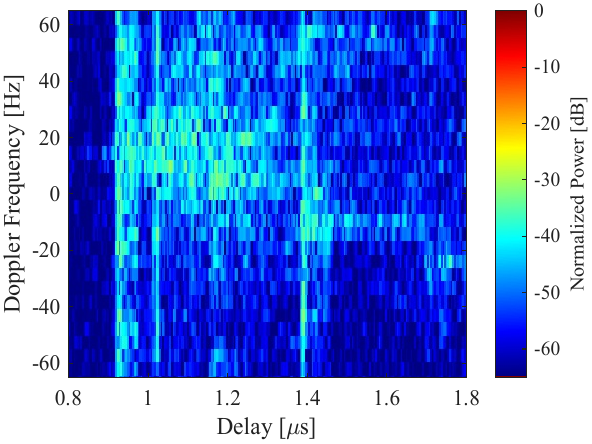}}
    \caption{Sparse multipath parameter estimation for a measured frame: (a) measured response; (b) reconstruction; and (c) residual.}
    \label{fig:dd_estimation_example}
\end{figure*}

The same procedure is then applied to the measured response. Fig.~\ref{fig:dd_estimation_example}(a), (b), and (c) show the measured power matrix over delay and Doppler, the reconstruction obtained from the estimated paths, and the residual, respectively. All three matrices use the maximum measured power as the common $0$~dB reference. The reconstruction contains the isolated high-power peaks, while background energy and delay-aligned diffuse structures remain in the residual. Because the measured multipath is dense, the finite extracted path set does not represent every component. The maximum residual power is $-30.28$~dB relative to the common reference. Only the extracted paths in $\widehat{\mathcal{G}}_{\mathrm{meas}}$, rather than the residual energy, are used to calculate the subsequent statistical parameters.

\section{Deterministic RT Branch}
\label{sec:rt_construction}

This section presents the construction of the 3D RT scene from satellite remote-sensing imagery, the classification of the resulting RT multipath, and the calibration of the electromagnetic material parameters.
\subsection{Remote-Sensing-Based 3D Reconstruction}

\begin{figure}[htb]
    \centering
    \includegraphics[width=0.8\columnwidth]{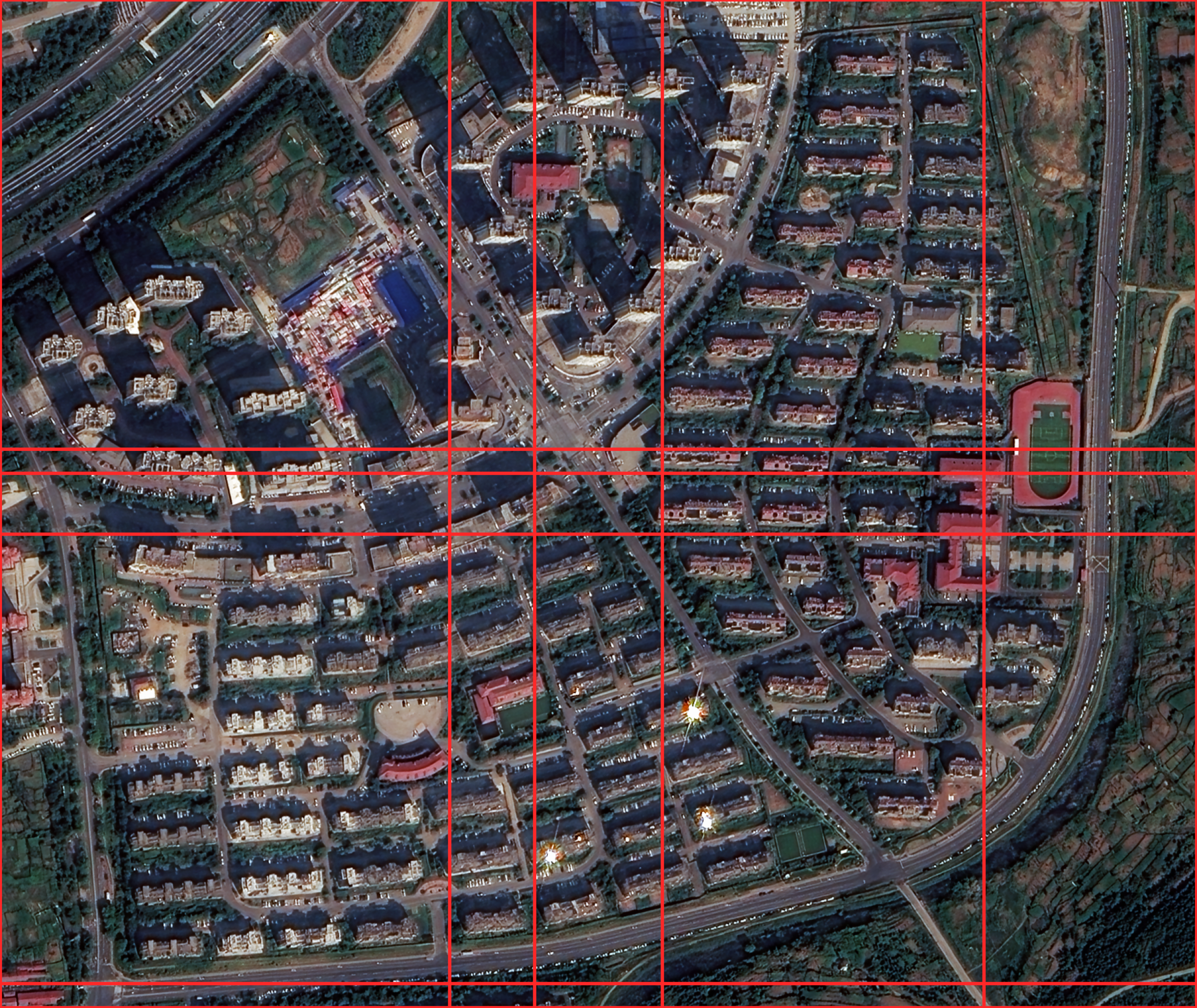}
    \caption{Satellite map and overlapping tiles for monocular depth estimation.}
    \label{fig:overlap_window}
\end{figure}

The ortho-rectified remote-sensing map has a ground sampling interval of 0.50 m but does not contain explicit height values. The Depth Anything V2 monocular depth foundation model~\cite{NEURIPS2024_26cfdcd8} is therefore used to estimate a relative depth map. The map is processed in overlapping tiles so that the full scene is not resized into one low-resolution model input. For the $k$th tile $\mathbf{I}_k$, the network prediction is written as
\begin{equation}
    \widehat{D}_k
    =
    \mathcal{M}_{\mathrm{DA}}(\mathbf{I}_k).
\end{equation}
In the reconstructed RT scene, the $1790\times1508$ pixel input map covers $895~\mathrm{m}\times754~\mathrm{m}$. The map is divided into $800\times800$ pixel tiles with a 190-pixel overlap, corresponding to a 610-pixel stride. The edge tiles are aligned with the image boundary, giving tile origins at $x=\{0,610,990\}$ and $y=\{0,610,708\}$, i.e., $3\times3=9$ depth-estimation tiles in total. Fig.~\ref{fig:overlap_window} shows the satellite map and the overlapping tile windows used for depth estimation. The shared regions between adjacent tiles provide common building areas for scale alignment and reduce discontinuities at tile boundaries.

\begin{figure}[htb]
    \centering
    \includegraphics[width=0.8\columnwidth]{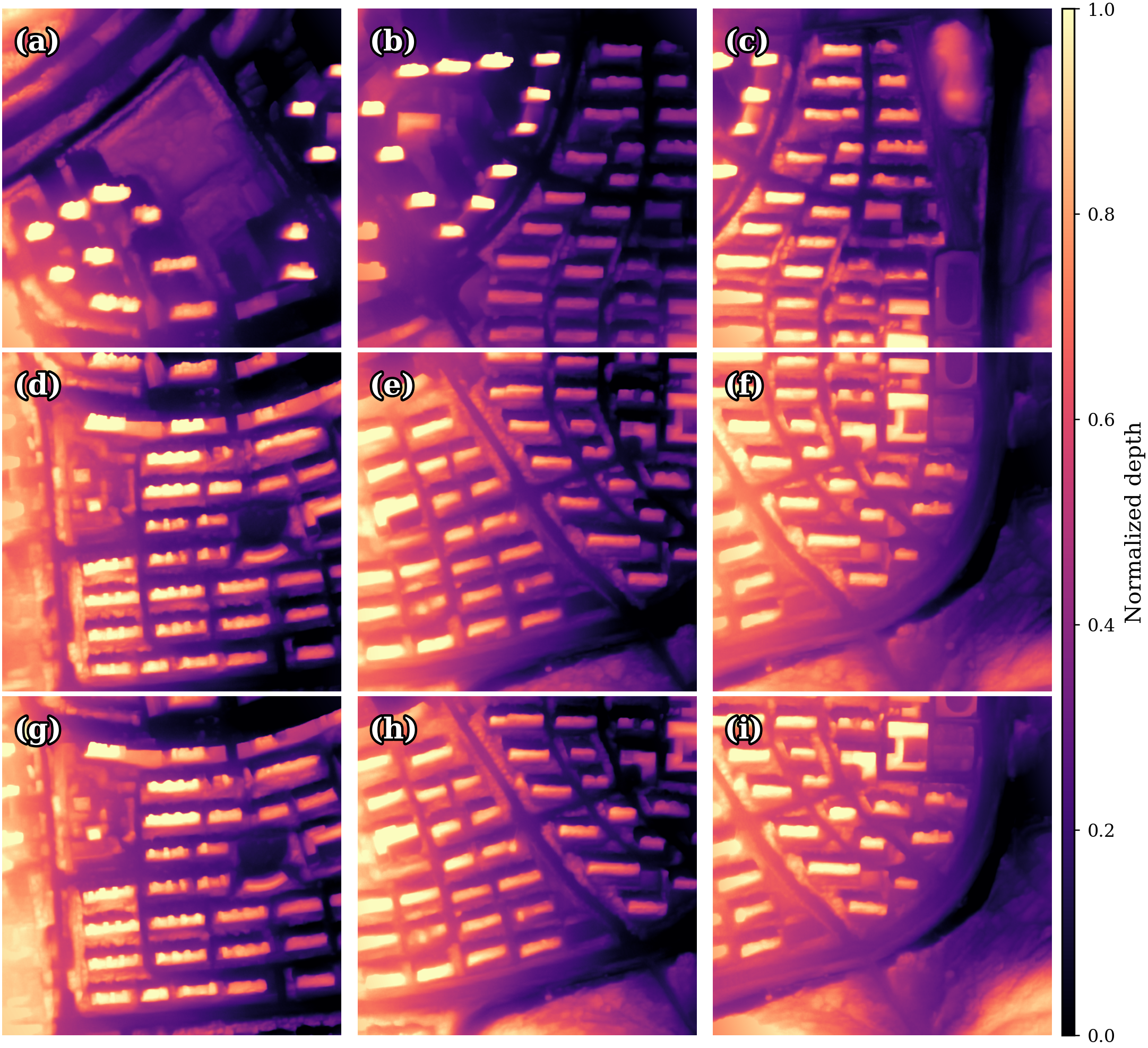}
    \caption{Tile-wise depth estimates after tilt correction and scale alignment.}
    \label{fig:tiled_depth_results}
\end{figure}

A full-map prediction is first computed and used as the scale reference for tile alignment. For each tile, pixels below a selected ground percentile are used as ground candidates. A planar depth tilt $\pi_k(u,v)$ is fitted on these samples and removed from the tile prediction. The corrected tile prediction is then linearly aligned to the corresponding patch of the full-map reference,
\begin{equation}
    \widetilde{D}_k(u,v)
    =
    \alpha_k\big(\widehat{D}_k(u,v)-\pi_k(u,v)\big)+\beta_k ,
\end{equation}
where $\alpha_k$ and $\beta_k$ are estimated by least-squares fitting. In regions already covered by previous tiles, an overlap-based linear alignment is applied before fusion. For pixels covered by multiple tiles, the mosaic value is computed as the cosine-weighted average of the aligned tile predictions. Fig.~\ref{fig:tiled_depth_results} shows the tile-wise depth estimates after local correction and scale alignment. The overlapping building regions are used to place the tile predictions on the same scale before fusion.

\begin{figure}[htb]
    \centering
    \begin{tabular}{@{}cc@{}}
        \includegraphics[height=0.3\columnwidth]{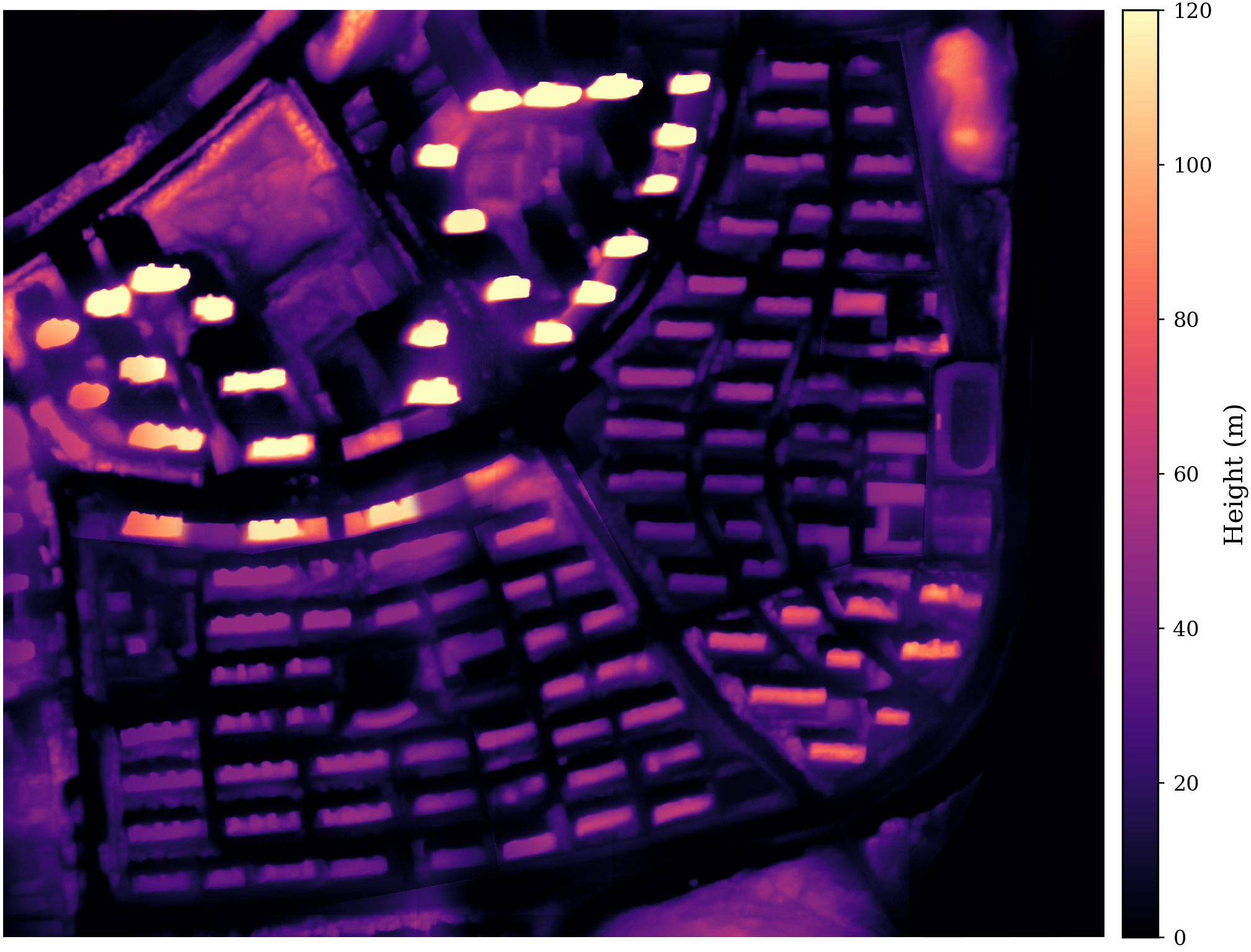} &
        \includegraphics[height=0.3\columnwidth]{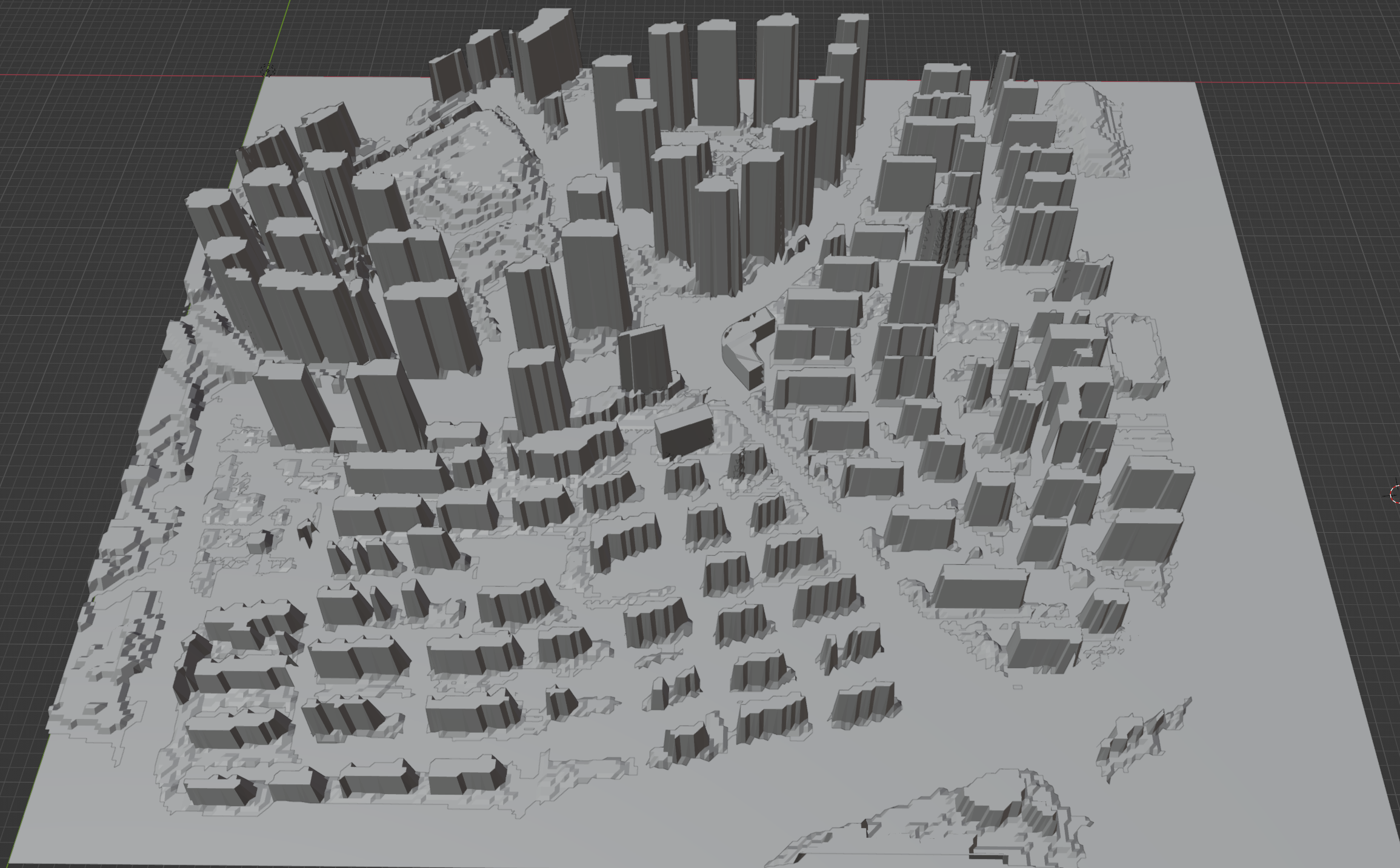} \\
        (a) & (b)
    \end{tabular}
    \caption{Remote-sensing-based 3D reconstruction: (a) estimated height map and (b) reconstructed RT scene.}
    \label{fig:height_blender_map}
\end{figure}

Let $\widetilde D[u,v]$ denote the fused relative-depth map after orienting its values such that larger values correspond to greater height. The relative values are linearly scaled to a metric height map by setting the maximum building height to $H_{\max}=120.00~\mathrm{m}$:
\begin{equation}
    h_{\mathrm m}[u,v]
    =
    H_{\max}
    \frac{
    \widetilde D[u,v]-D_{\min}
    }{
    D_{\max}-D_{\min}
    },
\end{equation}
where $D_{\min}=\min_{u,v}\widetilde D[u,v]$ and $D_{\max}=\max_{u,v}\widetilde D[u,v]$. The building height map $h_{\mathrm{b}}[u,v]$ is then obtained by setting heights below 5.00 m to zero:
\begin{equation}
    h_{\mathrm{b}}[u,v]
    =
    \begin{cases}
    h_{\mathrm{m}}[u,v], & h_{\mathrm{m}}[u,v]\geq 5.00~\mathrm{m},\\
    0, & h_{\mathrm{m}}[u,v]<5.00~\mathrm{m}.
    \end{cases}
\end{equation}
The remaining nonzero mask is split into connected components, and each component is treated as an isolated elevated object. To reduce the mesh complexity, components with a footprint area smaller than $A_{\min}=50.00~\mathrm{m}^2$ are removed, and the roofs of the retained components are flattened. For each retained component $\Omega_j$, let $\Omega_j^{\mathrm{top}}$ denote the pixels above a selected component-wise top-height quantile. The roof height is estimated as
\begin{equation}
    \bar{h}_j^{\mathrm{top}}
    =
    \frac{1}{|\Omega_j^{\mathrm{top}}|}
    \sum_{(u,v)\in\Omega_j^{\mathrm{top}}}
    h_{\mathrm{b}}[u,v].
\end{equation}
Pixels in $\Omega_j^{\mathrm{top}}$ are projected to $\bar{h}_j^{\mathrm{top}}$, while the remaining pixels in $\Omega_j$ are clipped to the wall-transition interval below the roof height. This component-wise regularization keeps adjacent buildings as separate objects and converts rounded monocular-depth roofs into flat RT surfaces. Fig.~\ref{fig:height_blender_map}(a) shows the resulting metric height map, and Fig.~\ref{fig:height_blender_map}(b) shows the corresponding Blender scene. The separated building components and flattened roofs in the height map are retained as individual objects and planar surfaces in the exported scene.

After regularization, the processed height map is converted into a triangular mesh on the 0.50 m grid. Each retained connected component is exported as an independent object, and a ground plane covering the full map is added as a separate object. The resulting OBJ scene is imported into Blender, where each object is assigned an initial material name. The Mitsuba Blender plugin~\cite{mitsuba_blender} is then used to convert the scene into the Extensible Markup Language (XML) format required by Sionna RT.

\subsection{RT Path Classification}

The RT channel is represented by $N_{\mathrm{RT}}$ paths with coefficient, delay, Doppler shift, and LoS indicator $\{a_i^{\mathrm{RT}},\tau_i^{\mathrm{RT}},\nu_i^{\mathrm{RT}},\ell_i\}_{i=1}^{N_{\mathrm{RT}}}$. The indicator $\ell_i\in\{0,1\}$ is directly provided by the RT output according to the path interaction count. The RT path power is $P_i^{\mathrm{RT}}=|a_i^{\mathrm{RT}}|^2$, and the total RT received power is $P_{\mathrm{tot}}=\sum_{i=1}^{N_{\mathrm{RT}}}P_i^{\mathrm{RT}}$.
If the RT result contains a LoS path, its index is denoted by $i_{\mathrm{L}}$, with $\ell_{i_{\mathrm{L}}}=1$. The LoS component contains only this direct RT path, and the LoS reference delay and Doppler are $\tau_{\mathrm{L}}=\tau_{i_{\mathrm{L}}}^{\mathrm{RT}}$ and $\nu_{\mathrm{L}}=\nu_{i_{\mathrm{L}}}^{\mathrm{RT}}$. For a LoS link, the excess delay of path $i$ is defined as $\Delta\tau_i=\tau_i^{\mathrm{RT}}-\tau_{\mathrm{L}}$.
If no RT path is labeled as LoS, $i_{\mathrm{L}}$ does not exist, and the LoS and LoS-tail components are not constructed for that link. The LoS-tail delay threshold $\tau_{\mathrm{T}}$ is used only for LoS links: non-LoS RT paths with $0<\Delta\tau_i\leq \tau_{\mathrm{T}}$ are treated as RT-resolved tail candidates, while the remaining non-LoS RT paths are assigned to the residual-NLoS component. Let $\boldsymbol{\chi}_i=(\chi_i^{\mathrm{L}},\chi_i^{\mathrm{T,RT}},\chi_i^{\mathrm{N}})$ denote the component assignment of path $i$:
\begin{equation}
\boldsymbol{\chi}_i=
\begin{cases}
(1,0,0), & \ell_i=1,\\
(0,1,0), & i_{\mathrm{L}}\ \text{exists},\ \ell_i=0,\ 0<\Delta\tau_i\leq \tau_{\mathrm{T}},\\
(0,0,1), & \text{otherwise}.
\end{cases}
\label{eq:rt_path_classification}
\end{equation}
The three entries correspond to the LoS component, the RT-resolved LoS-tail component, and the NLoS component, respectively.

\subsection{RT Electromagnetic Parameter Calibration}

After geometric reconstruction, the RT scene contains a ground plane and $N_{\mathrm{obj}}$ reconstructed object components. This subsection estimates the electromagnetic parameters of the reconstructed objects by minimizing the error between the measured and Sionna RT path losses.

For object $j=1,\ldots,N_{\mathrm{obj}}$, the trainable parameters are the relative permittivity $\varepsilon_{r,j}$ and conductivity $\sigma_j$. Every object is initialized with $\varepsilon_{r,j}^{(0)}=4$ and $\sigma_j^{(0)}=0.10~\mathrm{S/m}$. The scene geometry, ground-Rx position, and antenna configuration remain fixed. The Tx positions are taken from the measured links in each mini-batch and are not optimization variables.

For each mini-batch $\mathcal{B}$, $|\mathcal{B}|=16$ Tx positions are evaluated by Sionna RT. The solver uses a maximum interaction depth of two, with specular reflection and diffraction enabled. For link $m\in\mathcal{B}$, let $\mathcal{S}_m$ contain all valid RT paths and let $a_{m,i}^{\mathrm{RT}}$ denote the complex coefficient of path $i$. Under unit Tx power, the total RT channel gain $P_m^{\mathrm{RT}}$ and path loss $L_m^{\mathrm{RT}}$ are
\begin{equation}
\begin{aligned}
    P_m^{\mathrm{RT}}
    &=
    \sum_{i\in\mathcal{S}_m}
    \left|a_{m,i}^{\mathrm{RT}}\right|^2,\\
    L_m^{\mathrm{RT}}
    &=
    -10\log_{10} P_m^{\mathrm{RT}}.
\end{aligned}
\end{equation}

For calibration link $m$, let $\mathcal I_m$ denote its extracted-path index set. The measured total path loss calculated from the extracted path coefficients is $L_m^{\mathrm{meas}}=-10\log_{10}\sum_{p\in\mathcal I_m}|\widehat a_{m,p}^{\mathrm{meas}}|^2$. The calibration loss is the batch RMSE
\begin{equation}
    \mathcal{L}_{\mathcal{B}}
    =
    \sqrt{
    \frac{1}{|\mathcal{B}|}
    \sum_{m\in\mathcal{B}}
    \left(
    L_m^{\mathrm{RT}}-L_m^{\mathrm{meas}}
    \right)^2
    } .
\end{equation}
The same total path loss definition is used for both RT-LoS and RT-NLoS links.

At each iteration, the differentiable RT solver computes the path coefficients for one mini-batch, and the gradient of $\mathcal{L}_{\mathcal{B}}$ is back-propagated to $\{\varepsilon_{r,j},\sigma_j\}_{j=1}^{N_{\mathrm{obj}}}$. Adam runs for 200 iterations with a learning rate of $0.04$. The parameter state with the lowest RMSE over the calibration links is retained as the calibrated material set.

\begin{figure}[htbp]
    \centering
    \includegraphics[width=0.8\columnwidth]{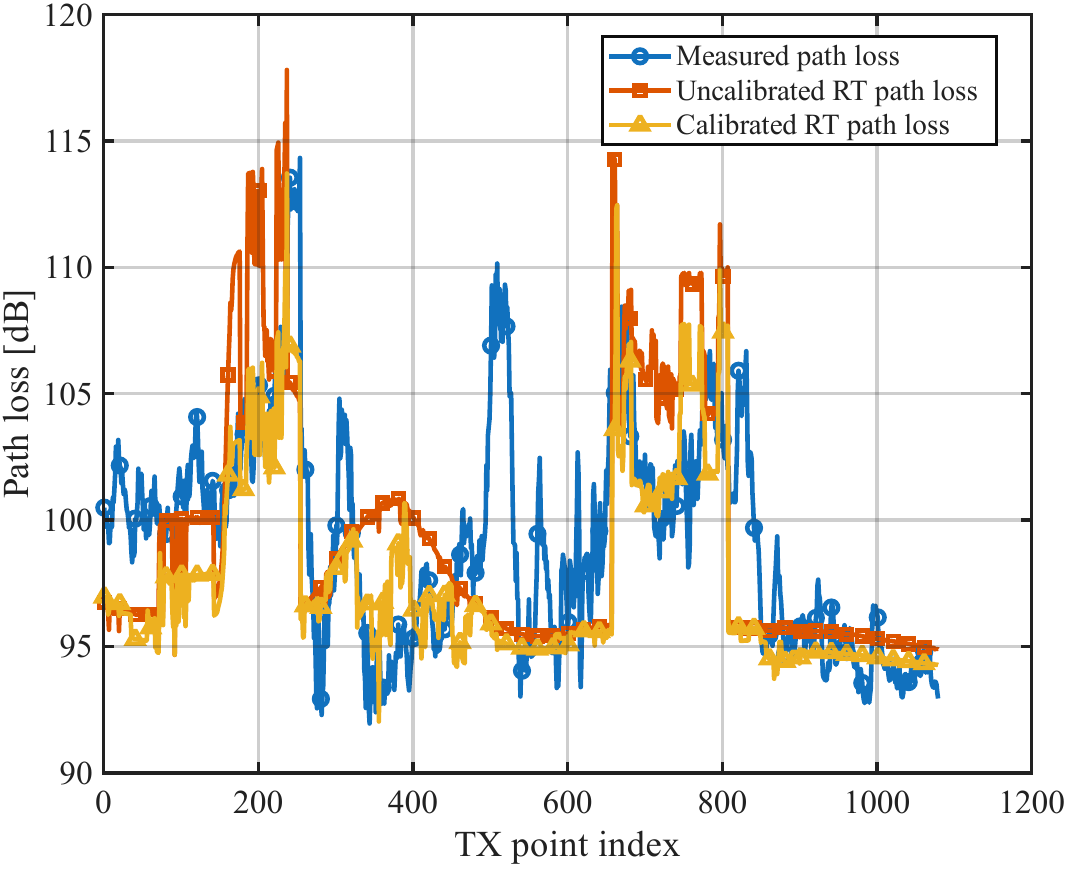}
    \caption{Measured and Sionna RT path loss before and after electromagnetic calibration.}
    \label{fig:emcal_path_loss}
\end{figure}

Fig.~\ref{fig:emcal_path_loss} summarizes the path loss calibration results. Over all matched links, the RMSE decreases by $1.10$~dB, from $5.45$~dB to $4.35$~dB, corresponding to a $20.18\%$ reduction. For the 832 LoS samples, the reduction is $0.05$~dB ($1.22\%$), from $4.10$~dB to $4.05$~dB. For the 247 NLoS samples, it is $2.78$~dB ($47.04\%$), from $5.91$~dB to $3.13$~dB. The zero-interaction direct path does not depend on the reconstructed-object parameters. Material calibration changes only the reflected and diffracted contributions, so its effect on the LoS path loss is limited.

\begin{figure}[htbp]
    \centering
    \includegraphics[width=0.8\columnwidth]{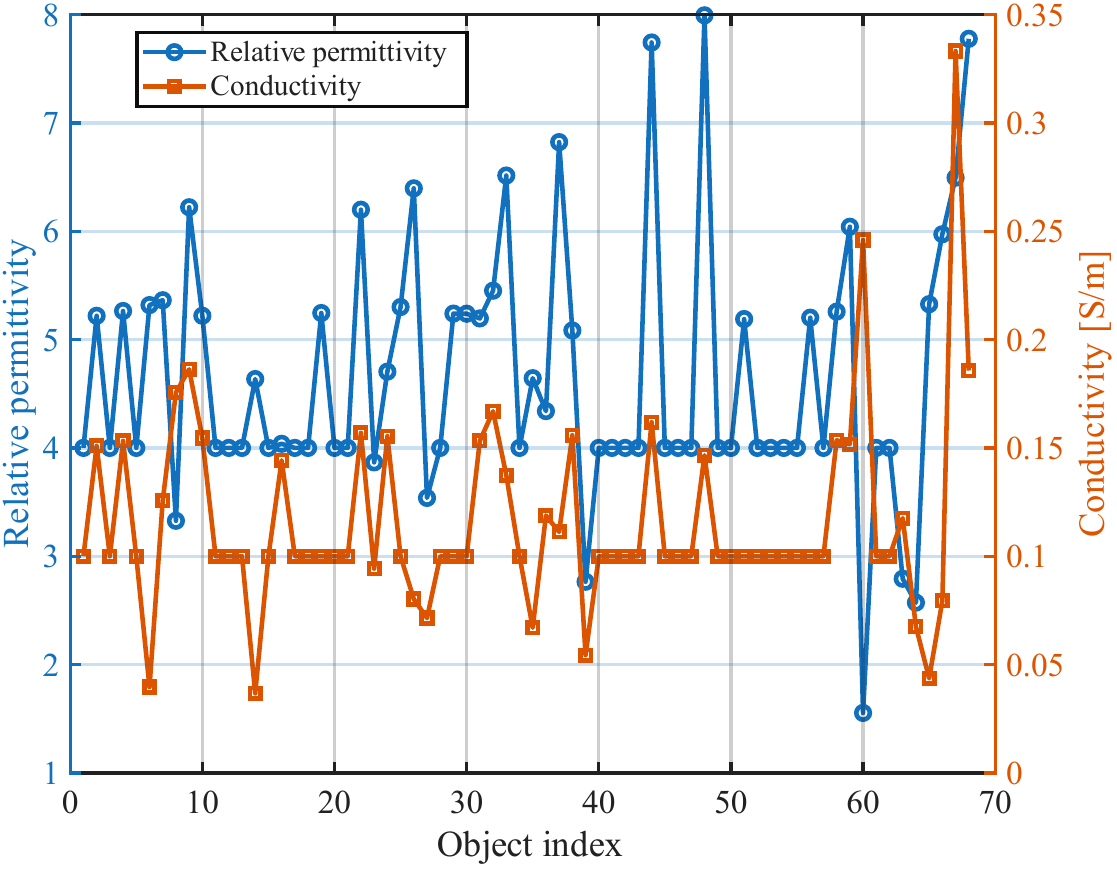}
    \caption{Calibrated relative permittivity and conductivity of the reconstructed RT objects.}
    \label{fig:emcal_materials}
\end{figure}

Fig.~\ref{fig:emcal_materials} shows that the calibrated relative permittivities range from approximately $1.60$ to $8.00$, while the conductivities range from approximately $0.04$ to $0.33~\mathrm{S/m}$. Because each reconstructed object may include walls, windows, and roof surfaces, these values are interpreted as effective object-level parameters rather than parameters of specific material types.

\section{Measurement-Statistical Branch}
\label{sec:statistical_modeling}

This section presents the LoS-link and NLoS-link parameter groups derived from the extracted multipath components, together with their fitted marginal distributions and Gaussian copula dependence models.
\subsection{Marginal Distributions of LoS-Link Parameters}

The LoS/NLoS label of each measured frame is provided by the RT result at the Tx position obtained from the RTK timestamp match and the fixed Rx position. Because this label is determined from the reconstructed static scene, it cannot fully represent transient blockage by local objects. Consequently, the RT-LoS measurement group can include abrupt changes in path power and dispersion caused by short-duration or partial blockage. These measured variations are retained when fitting the LoS-link parameter distributions.

For each measured frame, let $\mathcal I=\{1,\ldots,N_{\mathrm{est}}\}$ denote the extracted-path indices and $\widehat P_p=|\widehat a_p^{\mathrm{meas}}|^2$ denote the power of path $p$. For an RT-LoS frame, the extracted path whose delay is closest to the RT direct-path delay is used as the measured LoS reference, i.e., $p_{\mathrm L}=\arg\min_{p\in\mathcal I}|\widehat\tau_p^{\mathrm{meas}}-\tau_{\mathrm L}|$. For the measurement-based statistical modeling, the LoS-tail threshold introduced in Section~\ref{sec:channel_model} is set to $\tau_{\mathrm T}=100.00~\mathrm{ns}$. The remaining paths are divided into
\begin{equation}
\begin{aligned}
    \widehat{\mathcal S}_{\mathrm T}
    &=
    \left\{
    p\in\mathcal I:
    0<
    \widehat\tau_p^{\mathrm{meas}}
    -
    \widehat\tau_{p_{\mathrm L}}^{\mathrm{meas}}
    \leq\tau_{\mathrm T}
    \right\},\\
    \widehat{\mathcal S}_{\mathrm N}
    &=
    \mathcal I\setminus
    \left(
    \{p_{\mathrm L}\}\cup\widehat{\mathcal S}_{\mathrm T}
    \right).
\end{aligned}
\end{equation}
For any nonempty path set $\mathcal S$, let $\widehat P_{\mathcal S}=\sum_{p\in\mathcal S}\widehat P_p$ and $\overline{x}_{\mathcal S}=\sum_{p\in\mathcal S}\widehat P_p x_p/\widehat P_{\mathcal S}$. Its power-weighted RMS spread is
\begin{equation}
    \operatorname{rms}_{\mathcal S}(x)
    =
    \left[
    \frac{1}{\widehat P_{\mathcal S}}
    \sum_{p\in\mathcal S}\widehat P_p
    (x_p-\overline{x}_{\mathcal S})^2
    \right]^{1/2}.
    \label{eq:measured_rms_spread}
\end{equation}
For an RT-LoS frame, the component power ratios and LoS-tail path count are
\begingroup
\setlength{\abovedisplayskip}{6pt}
\setlength{\belowdisplayskip}{6pt}
\setlength{\abovedisplayshortskip}{6pt}
\setlength{\belowdisplayshortskip}{6pt}
\begin{equation}
\begin{aligned}
    \eta_{\mathrm T}
    &=
    \frac{\widehat P_{\widehat{\mathcal S}_{\mathrm T}}}
    {\widehat P_{p_{\mathrm L}}+
    \widehat P_{\widehat{\mathcal S}_{\mathrm T}}},
    &
    \xi_{\mathrm N}
    &=
    \frac{\widehat P_{\widehat{\mathcal S}_{\mathrm N}}}
    {\sum_{p\in\mathcal I}\widehat P_p},\\
    N_{\mathrm T}
    &=
    \left|\widehat{\mathcal S}_{\mathrm T}\right|.
\end{aligned}
\end{equation}
\endgroup
For each component set, the delay spread is obtained by applying $\operatorname{rms}_{\mathcal S}(\cdot)$ to $\widehat\tau^{\mathrm{meas}}$, and the normalized Doppler spread is obtained by applying it to $\widehat\nu^{\mathrm{meas}}$ and dividing by $f_{\max}$. Using $\widehat{\mathcal S}_{\mathrm T}$ gives $\sigma_{\tau,\mathrm T}$ and $\kappa_{\nu,\mathrm T}$, while using $\widehat{\mathcal S}_{\mathrm N}$ gives $\sigma_{\tau,\mathrm N}$ and $\kappa_{\nu,\mathrm N}$. Frames with an empty component are excluded from the fit of the parameters that require that component. The marginal distributions are fitted separately using maximum likelihood estimation (MLE)~\cite{myung2003mle}.

For each measured frame with a valid LoS reference, two groups of statistics are extracted. The LoS-tail group contains the tail power ratio, RMS delay spread, normalized Doppler spread, and path count. The residual-NLoS group contains the residual power ratio, RMS delay spread, and normalized Doppler spread after removing the LoS and LoS-tail paths. Because the extraction uses $N_{\max}=60$ and a 40.00~dB relative-power threshold, $N_{\mathrm T}$ represents the number of detectable LoS-tail paths under this extraction configuration rather than the total number of physical propagation paths. For the zero-truncated negative binomial fit of $N_{\mathrm T}$, $r_{\mathrm{NB}}$ and $p_{\mathrm{NB}}$ denote the shape and success-probability parameters under the convention $\Pr(N_{\mathrm T}=n)\propto\binom{n+r_{\mathrm{NB}}-1}{n}(1-p_{\mathrm{NB}})^n p_{\mathrm{NB}}^{r_{\mathrm{NB}}}$ for integers $n\geq1$.

Table~\ref{tab:los_statistical_parameters} summarizes the fitted distributions for the LoS-tail and residual-NLoS parameter groups. For the LoS-tail group, $\eta_{\mathrm T}$ follows a Beta distribution with $\alpha=2.16$ and $\beta=5.90$; $\sigma_{\tau,\mathrm T}$ and $\kappa_{\nu,\mathrm T}$ follow Weibull distributions with scale--shape pairs $(18.81~\mathrm{ns},1.72)$ and $(0.02,1.79)$, respectively; and $N_{\mathrm T}$ follows a zero-truncated negative binomial distribution with $r_{\mathrm{NB}}=5.57$ and $p_{\mathrm{NB}}=0.26$. For the residual-NLoS group, $\xi_{\mathrm N}$ follows a Beta distribution with $\alpha=0.83$ and $\beta=8.21$, while $\sigma_{\tau,\mathrm N}$ and $\kappa_{\nu,\mathrm N}$ follow Weibull distributions with scale--shape pairs $(133.80~\mathrm{ns},1.79)$ and $(0.09,1.28)$, respectively.
\begin{table}[H]
\centering
\caption{Fitted statistical distributions of the LoS-link parameters.}
\label{tab:los_statistical_parameters}
\footnotesize
\setlength{\tabcolsep}{2.5pt}
\begin{tabular*}{\columnwidth}{@{\extracolsep{\fill}}lll@{}}
\toprule
Parameter & Distribution & Fitted parameters \\
\midrule
\multicolumn{3}{l}{\textit{LoS-tail}} \\
$\eta_{\mathrm{T}}$ & Beta & $\alpha=2.16,\ \beta=5.90$ \\
$\sigma_{\tau,\mathrm{T}}$ & Weibull & scale $=18.81$ ns, shape $=1.72$ \\
$\kappa_{\nu,\mathrm{T}}$ & Weibull & scale $=0.02$, shape $=1.79$ \\
$N_{\mathrm{T}}$ & \shortstack[l]{Zero-truncated\\negative binomial} & $r_{\mathrm{NB}}=5.57,\ p_{\mathrm{NB}}=0.26$ \\
\midrule
\multicolumn{3}{l}{\textit{Residual-NLoS}} \\
$\xi_{\mathrm{N}}$ & Beta & $\alpha=0.83,\ \beta=8.21$ \\
$\sigma_{\tau,\mathrm{N}}$ & Weibull & scale $=133.80$ ns, shape $=1.79$ \\
$\kappa_{\nu,\mathrm{N}}$ & Weibull & scale $=0.09$, shape $=1.28$ \\
\bottomrule
\end{tabular*}
\end{table}

\begin{figure}[!t]
    \centering
    \subfloat[$\eta_{\mathrm{T}}$]{\includegraphics[width=0.48\columnwidth]{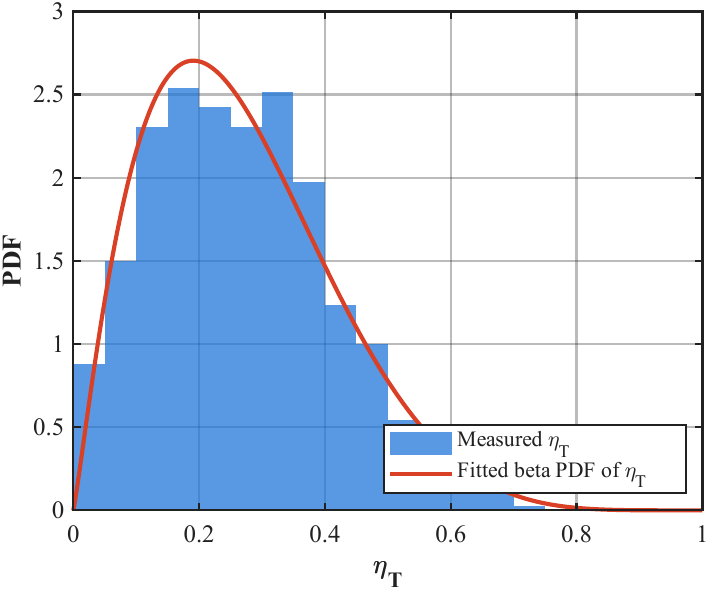}}
    \hfill
    \subfloat[$\sigma_{\tau,\mathrm{T}}$]{\includegraphics[width=0.48\columnwidth]{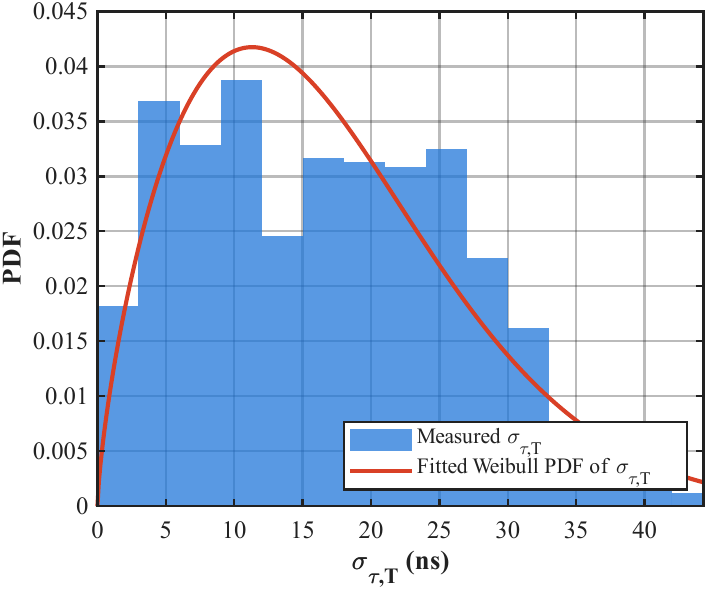}}\\[0.6em]
    \subfloat[$\kappa_{\nu,\mathrm{T}}$]{\includegraphics[width=0.48\columnwidth]{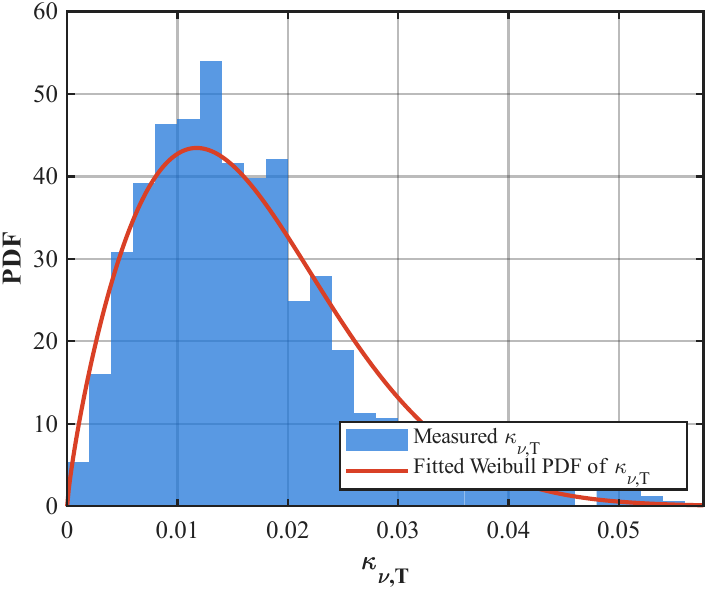}}
    \hfill
    \subfloat[$N_{\mathrm{T}}$]{\includegraphics[width=0.48\columnwidth]{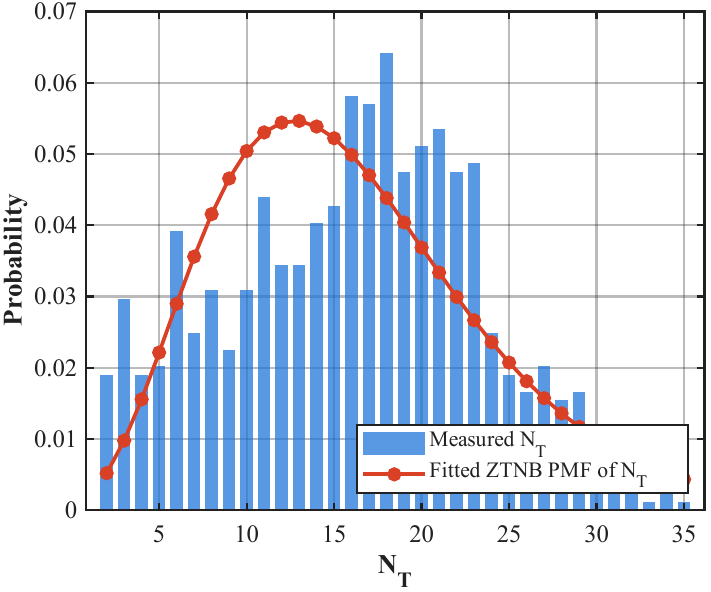}}
    \caption{Marginal fits of the LoS-tail parameters.}
    \label{fig:los_tail_marginals}
\end{figure}

Fig.~\ref{fig:los_tail_marginals} compares the measured histograms and fitted marginal distributions of $\boldsymbol{\Theta}_{\mathrm{T}}=[\eta_{\mathrm{T}},\sigma_{\tau,\mathrm{T}},\kappa_{\nu,\mathrm{T}},N_{\mathrm{T}}]^{\mathrm T}$. The fitted curves for $\eta_{\mathrm T}$, $\sigma_{\tau,\mathrm T}$, and $\kappa_{\nu,\mathrm T}$ peak at approximately $0.20$, $10$~ns, and $0.01$, respectively, near the largest measured histogram bars. The fitted distribution of $N_{\mathrm T}$ peaks at approximately 13 paths, while the largest measured bars occur at approximately 17--18 paths.

\begin{figure}[!t]
    \centering
    \subfloat[$\xi_{\mathrm{N}}$]{\includegraphics[width=0.48\columnwidth]{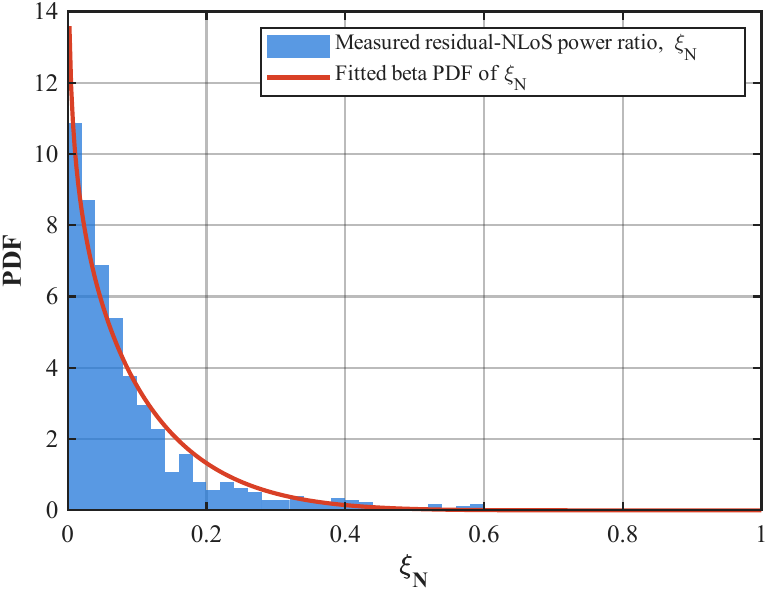}}
    \hfill
    \subfloat[$\kappa_{\nu,\mathrm{N}}$]{\includegraphics[width=0.48\columnwidth]{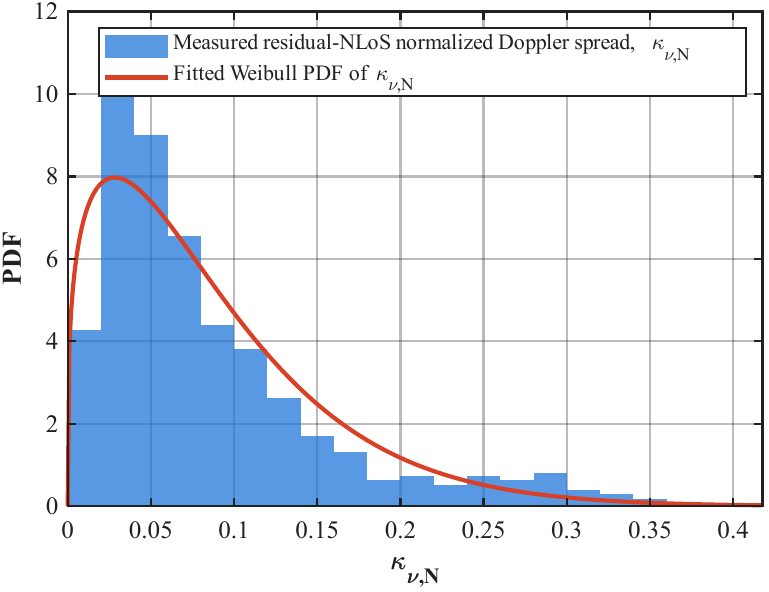}}\\[0.6em]
    \subfloat[$\sigma_{\tau,\mathrm{N}}$]{\includegraphics[width=0.48\columnwidth]{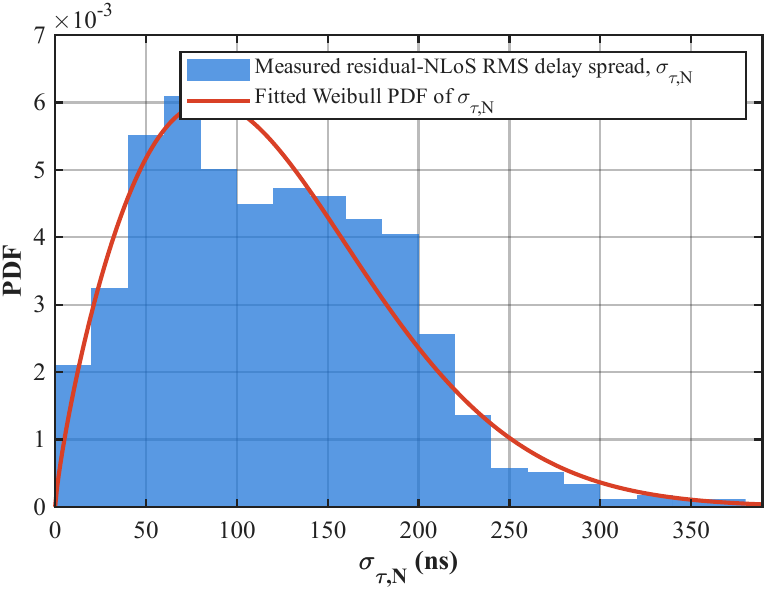}}
    \caption{Marginal fits of the LoS-link residual-NLoS parameters.}
    \label{fig:nlos_marginals}
\end{figure}

Fig.~\ref{fig:nlos_marginals} compares the measured histograms and fitted marginal distributions of the LoS-link residual-NLoS parameters. The fitted $\xi_{\mathrm N}$ density is largest near zero, while the fitted $\sigma_{\tau,\mathrm N}$ and $\kappa_{\nu,\mathrm N}$ densities peak at approximately $85$~ns and $0.03$, respectively. The fitted curves differ from the histograms mainly in the first bins of $\xi_{\mathrm N}$ and at the largest measured values of the two spread parameters.

\subsection{Marginal Distributions of NLoS-Link Parameters}

For measured frames identified as NLoS, no LoS reference is available and the LoS-tail power split is not defined. All extracted paths form the NLoS set $\widehat{\mathcal S}_{\mathrm N}^{\mathrm{NLoS}}=\mathcal I$. Applying \eqref{eq:measured_rms_spread} to their delays and Doppler shifts gives $\sigma_{\tau,\mathrm N}^{\mathrm{NLoS}}$ and $\kappa_{\nu,\mathrm N}^{\mathrm{NLoS}}$, respectively. These parameters form $\boldsymbol{\Theta}_{\mathrm{N}}^{\mathrm{NLoS}}=[\sigma_{\tau,\mathrm{N}}^{\mathrm{NLoS}},\kappa_{\nu,\mathrm{N}}^{\mathrm{NLoS}}]^{\mathrm T}$ and are fitted by MLE. As summarized in Table~\ref{tab:nlos_statistical_parameters}, $\sigma_{\tau,\mathrm N}^{\mathrm{NLoS}}$ follows a Weibull distribution with scale $139.99$~ns and shape $3.32$, while $\kappa_{\nu,\mathrm N}^{\mathrm{NLoS}}$ follows a Weibull distribution with scale $0.05$ and shape $2.73$.

\begin{table}[H]
\centering
\caption{Fitted NLoS-link parameter distributions.}
\label{tab:nlos_statistical_parameters}
\begin{tabular*}{0.94\columnwidth}{@{\extracolsep{\fill}}ccl@{}}
\toprule
Parameter & Distribution & Fitted parameters \\
\midrule
$\sigma_{\tau,\mathrm{N}}^{\mathrm{NLoS}}$ & Weibull & scale $=139.99$ ns, shape $=3.32$ \\
$\kappa_{\nu,\mathrm{N}}^{\mathrm{NLoS}}$ & Weibull & scale $=0.05$, shape $=2.73$ \\
\bottomrule
\end{tabular*}
\end{table}

\begin{figure}[H]
    \centering
    \subfloat[$\sigma_{\tau,\mathrm{N}}^{\mathrm{NLoS}}$]{\includegraphics[width=0.48\columnwidth]{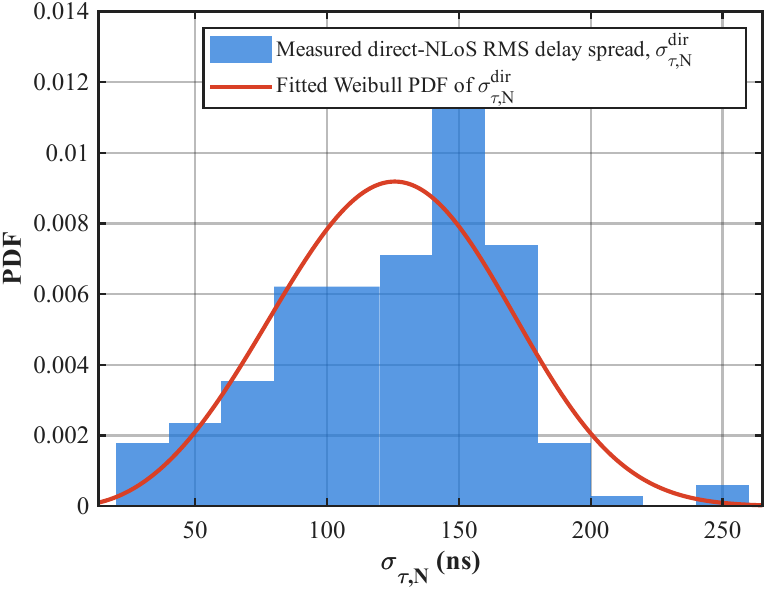}}
    \hfill
    \subfloat[$\kappa_{\nu,\mathrm{N}}^{\mathrm{NLoS}}$]{\includegraphics[width=0.48\columnwidth]{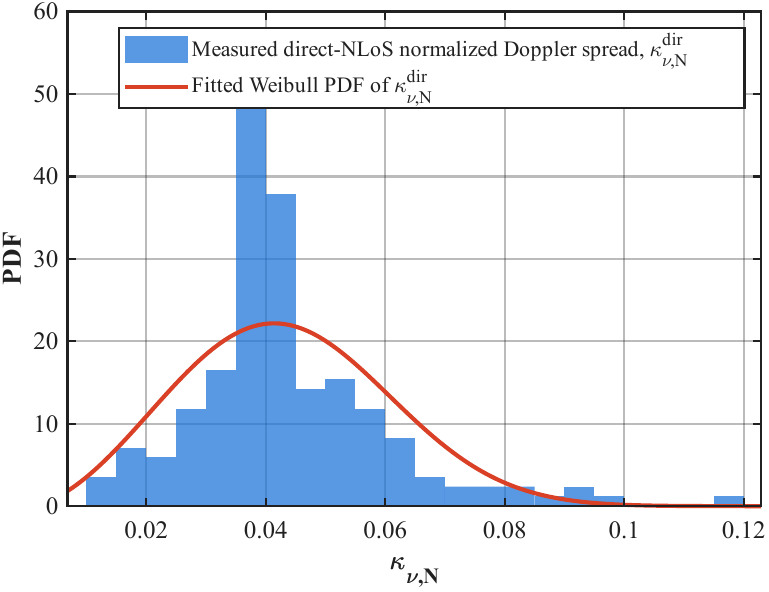}}
    \caption{Marginal fits of the NLoS-link parameters.}
    \label{fig:nlos_link_marginals}
\end{figure}

Fig.~\ref{fig:nlos_link_marginals} compares the measured histograms with the corresponding Weibull fits for the NLoS-link RMS delay spread and normalized Doppler spread. The fitted modes are $125.69$~ns and $0.04$, respectively, and are close to the largest measured histogram bars. Differences remain in several bins near the lower and upper ends of both histograms. These fitted distributions are used to sample the spread parameters for RT-NLoS links.

\subsection{Gaussian Copula Dependence Modeling}

The preceding marginal distributions describe each parameter separately. To model the correlations observed within each parameter group, a Gaussian copula is fitted to the LoS-tail vector $\boldsymbol{\Theta}_{\mathrm T}$, the LoS-link residual-NLoS vector $\boldsymbol{\Theta}_{\mathrm N}=[\xi_{\mathrm N},\sigma_{\tau,\mathrm N},\kappa_{\nu,\mathrm N}]^{\mathrm T}$, and the NLoS-link vector $\boldsymbol{\Theta}_{\mathrm N}^{\mathrm{NLoS}}$~\cite{song2000gaussian}.

For a generic parameter vector $\boldsymbol{\Theta}=[\theta_1,\ldots,\theta_D]^{\mathrm T}$, let $F_d$ denote the fitted marginal CDF of $\theta_d$. For a continuous parameter, $\widetilde F_d=F_d$. For the discrete tail count, the mid-distribution transform $\widetilde F_d(n)=F_d(n^-)+\Pr(N_{\mathrm T}=n)/2$ is used. Each measured sample is mapped to the standard Gaussian domain as
\begin{equation}
\begin{aligned}
    u_d
    &=
    \widetilde F_d(\theta_d),\\
    z_d
    &=
    \Phi^{-1}(u_d),
    \quad d=1,\ldots,D ,
\end{aligned}
\end{equation}
where $\Phi^{-1}(\cdot)$ is the inverse standard normal CDF. The mid-distribution transform handles the discrete values of $N_{\mathrm T}$. The correlation matrix $\mathbf R$ is estimated from the transformed vectors $\mathbf z=[z_1,\ldots,z_D]^{\mathrm T}$. The Gaussian copula and the joint distribution of $\boldsymbol{\Theta}$ are
\begin{equation}
\begin{aligned}
    C_{\mathbf R}(\mathbf u)
    &=
    \Phi_{\mathbf R}
    \left(
    \Phi^{-1}(u_1),\ldots,\Phi^{-1}(u_D)
    \right),\\
    F_{\boldsymbol{\Theta}}(\boldsymbol{\theta})
    &=
    C_{\mathbf R}
    \left(
    F_1(\theta_1),\ldots,
    F_D(\theta_D)
    \right),
\end{aligned}
\end{equation}
where $\Phi_{\mathbf R}(\cdot)$ is the multivariate Gaussian CDF with correlation matrix $\mathbf R$. During sampling, the inverse CDF of each marginal distribution converts the generated uniform samples into values of the corresponding model parameter.

\begin{figure}[htbp]
    \centering
    \includegraphics[width=\columnwidth]{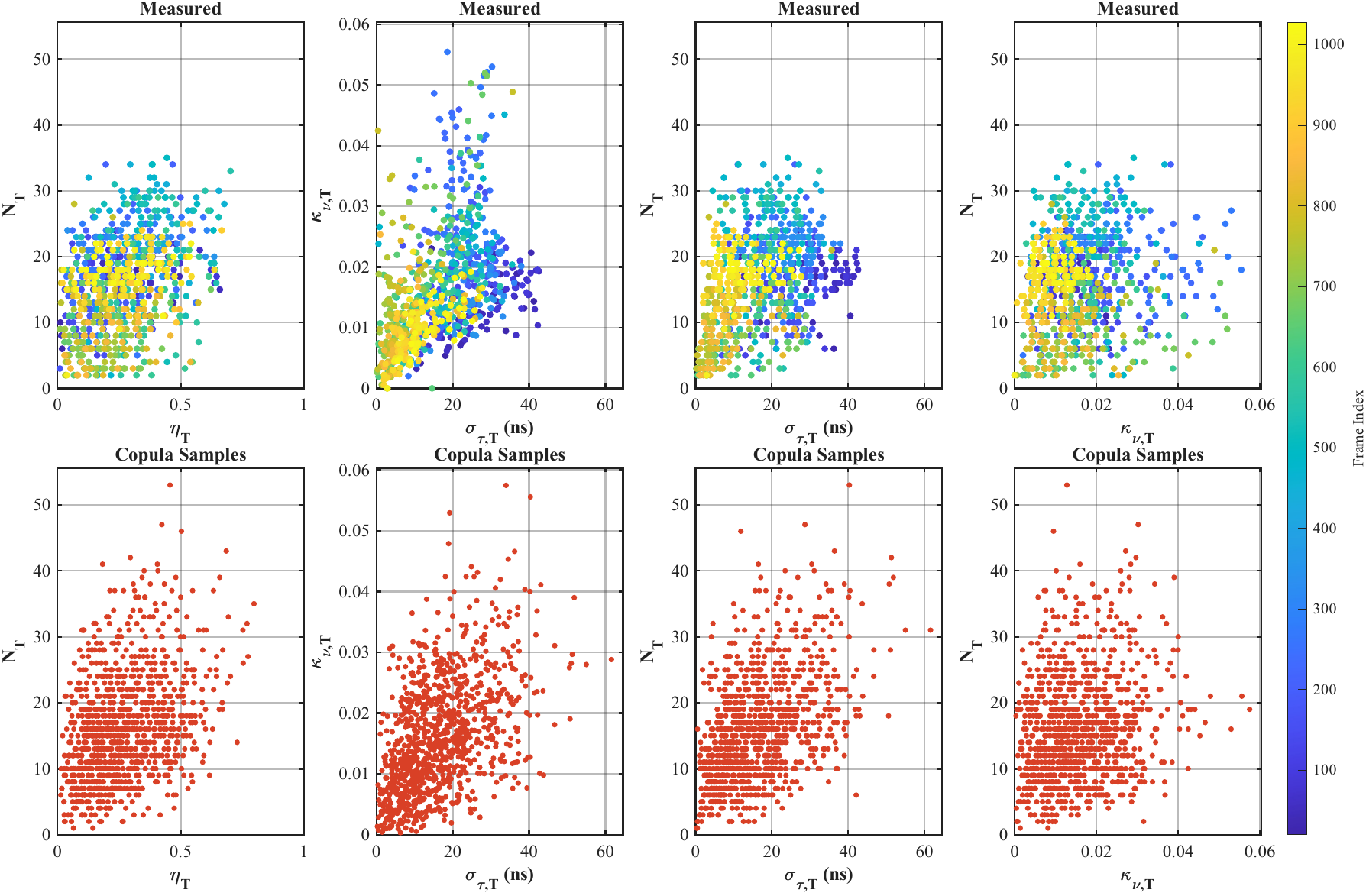}
    \caption{Pairwise measured and Gaussian copula samples for the retained LoS-tail parameter pairs.}
    \label{fig:los_tail_copula_corr}
\end{figure}

For the LoS-tail parameters, the fitted correlation matrix is
\begin{equation}
\mathbf{R}_{\mathrm{T}}=
\begin{bmatrix}
1.00 & 0.01 & -0.09 & 0.37\\
0.01 & 1.00 & 0.51 & 0.49\\
-0.09 & 0.51 & 1.00 & 0.22\\
0.37 & 0.49 & 0.22 & 1.00
\end{bmatrix}.
\end{equation}
The coefficients of $\eta_{\mathrm T}$ with $\sigma_{\tau,\mathrm T}$ and $\kappa_{\nu,\mathrm T}$ are $0.01$ and $-0.09$, respectively. The coefficient between $\sigma_{\tau,\mathrm T}$ and $\kappa_{\nu,\mathrm T}$ is $0.51$, while $N_{\mathrm T}$ has coefficients of $0.37$, $0.49$, and $0.22$ with $\eta_{\mathrm T}$, $\sigma_{\tau,\mathrm T}$, and $\kappa_{\nu,\mathrm T}$, respectively. Fig.~\ref{fig:los_tail_copula_corr} compares measured and copula-generated samples for the four displayed parameter pairs. The pairs $(\eta_{\mathrm T},\sigma_{\tau,\mathrm T})$ and $(\eta_{\mathrm T},\kappa_{\nu,\mathrm T})$ are omitted because their absolute coefficients are $0.01$ and $0.09$.

\begin{figure}[htbp]
    \centering
    \includegraphics[width=\columnwidth]{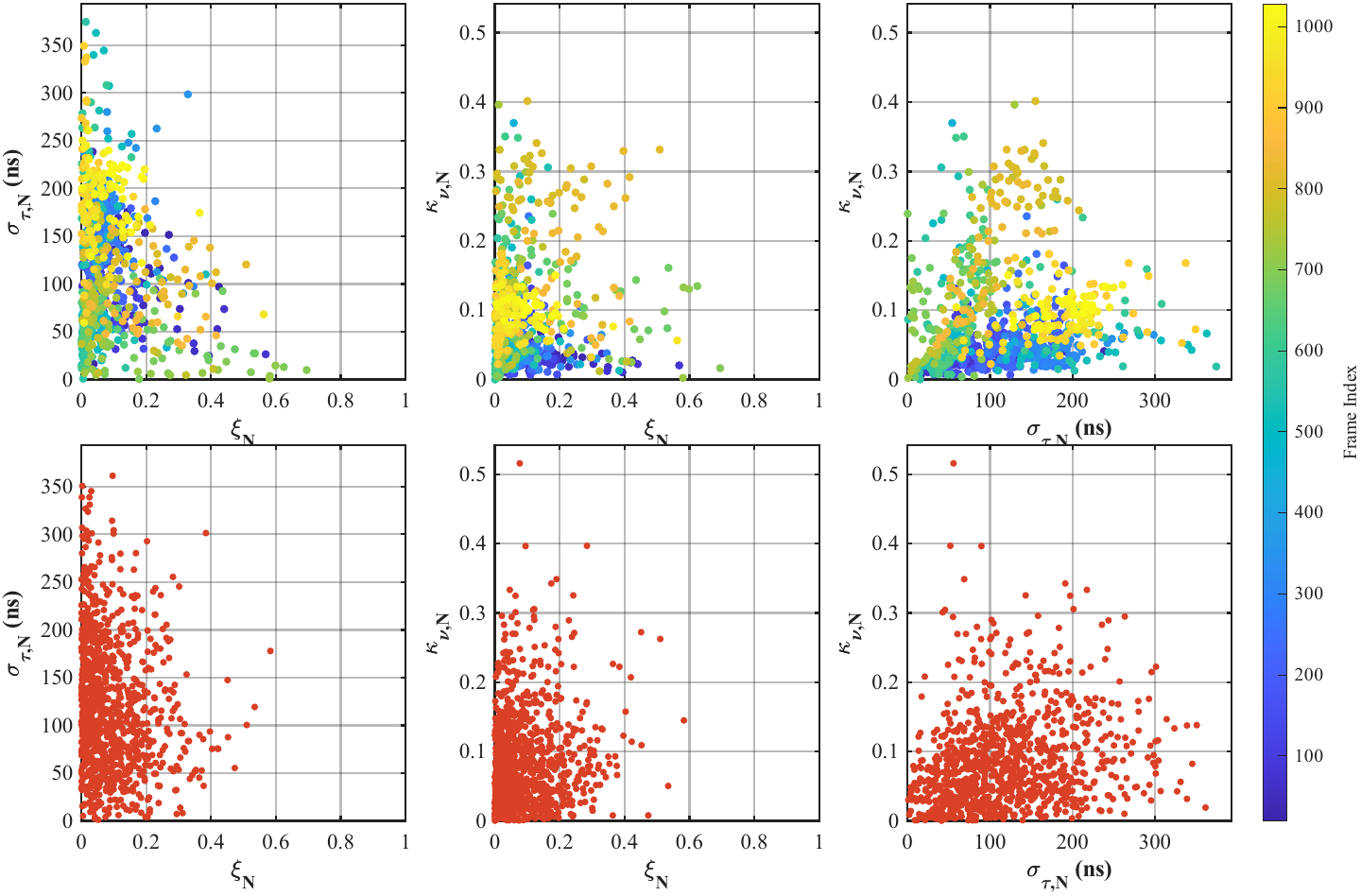}
    \caption{Pairwise measured and Gaussian copula samples for the LoS-link residual-NLoS parameters.}
    \label{fig:nlos_copula_corr}
\end{figure}

For the LoS-link residual-NLoS parameters, the fitted correlation matrix is
\begin{equation}
\mathbf{R}_{\mathrm{N}}=
\begin{bmatrix}
1.00 & -0.15 & 0.12\\
-0.15 & 1.00 & 0.18\\
0.12 & 0.18 & 1.00
\end{bmatrix}.
\end{equation}
The three coefficients have magnitudes below $0.20$: the coefficient between $\xi_{\mathrm N}$ and $\sigma_{\tau,\mathrm N}$ is $-0.15$, while those between $\kappa_{\nu,\mathrm N}$ and the other two parameters are $0.12$ and $0.18$. These values are lower than the $0.51$ coefficient between the LoS-tail delay and Doppler spreads and the $0.22$--$0.49$ coefficients involving $N_{\mathrm T}$. Accordingly, neither the measured nor the generated samples in Fig.~\ref{fig:nlos_copula_corr} show a clear linear trend.

\begin{figure}[htbp]
    \centering
    \includegraphics[width=\columnwidth]{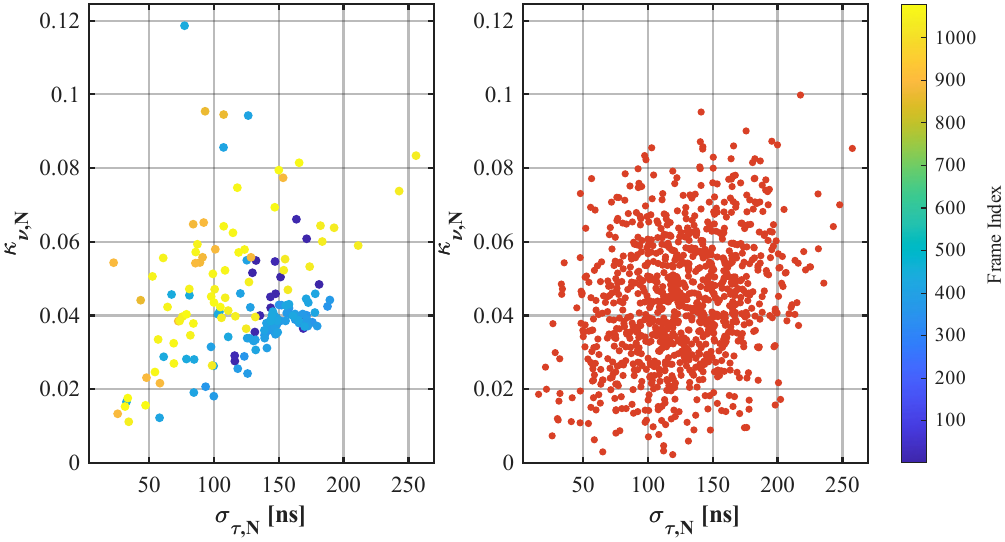}
    \caption{Measured and Gaussian copula samples for the NLoS-link parameters.}
    \label{fig:nlos_copula}
\end{figure}

For the NLoS-link parameters, the fitted correlation matrix is
\begin{equation}
\mathbf{R}_{\mathrm{N}}^{\mathrm{NLoS}}=
\begin{bmatrix}
1.00 & 0.29\\
0.29 & 1.00
\end{bmatrix}.
\end{equation}
The coefficient between the NLoS-link RMS delay spread and normalized Doppler spread is $0.29$. Fig.~\ref{fig:nlos_copula} shows a weak positive trend in both the measured and generated samples for this parameter pair. Each generated point is sampled independently from the fitted model and is not paired with the measured point at the same frame index.

\section{RT Augmentation Branch: Realization and Validation}
\label{sec:validation}

This section presents the NLoS power reallocation and LoS-tail augmentation methods, their numerical realization, and the validation of the resulting RS-ART channels against measurements and calibrated RT results.
\subsection{NLoS Power Reallocation}

The NLoS power reallocation retains the RT path delays, Doppler shifts, and phases while adapting the path powers to measurement-derived component statistics. For an RT-LoS link, it is applied to the residual-NLoS component outside the LoS-tail delay region; for an RT-NLoS link, the NLoS component represents the full channel. The corresponding statistical parameter vectors are
\begin{equation}
\begin{aligned}
    \boldsymbol{\Theta}_{\mathrm{N}}
    &=
    [\xi_{\mathrm{N}},\sigma_{\tau,\mathrm{N}},
    \kappa_{\nu,\mathrm{N}}]^{\mathrm T},\\
    \boldsymbol{\Theta}_{\mathrm{N}}^{\mathrm{NLoS}}
    &=
    [\sigma_{\tau,\mathrm{N}}^{\mathrm{NLoS}},
    \kappa_{\nu,\mathrm{N}}^{\mathrm{NLoS}}]^{\mathrm T}.
\end{aligned}
\label{eq:nlos_parameter_vectors}
\end{equation}
Here, $0\leq\xi_{\mathrm{N}}\leq1$ is the residual-NLoS power ratio for an RT-LoS link, $\sigma_{\tau,\mathrm{N}}>0$ is the RMS delay spread, and $\kappa_{\nu,\mathrm{N}}\geq0$ is the normalized Doppler spread. The physical Doppler spread target is $\sigma_{\nu,\mathrm{N}}=\kappa_{\nu,\mathrm{N}}f_{\max}$; the superscript ``NLoS'' denotes the corresponding targets for an RT-NLoS link.

The power allocated to the NLoS component depends on the RT link state, while the total RT received power is preserved:
\begin{equation}
\left(P_{\mathrm{N}},P_{\mathrm{L+T}}\right)
=
\begin{cases}
\left(\xi_{\mathrm{N}}P_{\mathrm{tot}},
      (1-\xi_{\mathrm{N}})P_{\mathrm{tot}}\right),
      & \text{RT-LoS},\\
\left(P_{\mathrm{tot}},0\right),
      & \text{RT-NLoS}
\end{cases}.
\label{eq:nlos_power_allocation}
\end{equation}
Thus, an RT-NLoS link has no LoS-related power split, whereas the remaining power $P_{\mathrm{L+T}}$ of an RT-LoS link is divided by the LoS-tail model.

Let $\mathcal{S}_{\mathrm{N}}=\{i:\chi_i^{\mathrm{N}}=1\}$ denote the retained NLoS RT paths. Their power transformation is represented by the operator
\begin{equation}
\{P_i^{\mathrm{N}}\}_{i\in\mathcal{S}_{\mathrm{N}}}
=
\mathcal{A}_{\mathrm{N}}
\left(
\{P_i^{\mathrm{RT}},\tau_i^{\mathrm{RT}},\nu_i^{\mathrm{RT}}\}_{i\in\mathcal{S}_{\mathrm{N}}};
P_{\mathrm{N}},\sigma_{\tau}^{\star},\sigma_{\nu}^{\star}
\right),
\label{eq:nlos_power_operator}
\end{equation}
where $(\sigma_{\tau}^{\star},\sigma_{\nu}^{\star})$ equals the residual-NLoS spread targets for an RT-LoS link and the NLoS spread targets for an RT-NLoS link. The operator produces nonnegative powers that sum to $P_{\mathrm{N}}$ and approach the specified RMS delay and Doppler spreads without changing the RT path locations or phases. Section~\ref{sec:channel_realization} gives its numerical realization.

\subsection{LoS-Tail Component Modeling}

The LoS-tail model describes short-delay multipath relative to the RT LoS path. Its statistical parameter vector is
\begin{equation}
    \boldsymbol{\Theta}_{\mathrm{T}}
    =
    [\eta_{\mathrm{T}},\sigma_{\tau,\mathrm{T}},
    \kappa_{\nu,\mathrm{T}},N_{\mathrm{T}}]^{\mathrm T},
\label{eq:los_tail_parameter_vector}
\end{equation}
where $0\leq\eta_{\mathrm{T}}\leq1$ is the tail power ratio within the LoS-plus-tail component, $\sigma_{\tau,\mathrm{T}}>0$ is the power-weighted RMS delay spread of the LoS-tail paths, $\kappa_{\nu,\mathrm{T}}\geq0$ is the normalized Doppler spread, and $N_{\mathrm{T}}\in\mathbb N^+$ is the sampled LoS-tail path-count target used to determine whether additional paths are required. Because the fitted count is obtained from an extraction with maximum count $N_{\max}$, samples above $N_{\max}$ are set to $N_{\max}$ during channel realization. The corresponding physical Doppler spread is $\sigma_{\nu,\mathrm{T}}=\kappa_{\nu,\mathrm{T}}f_{\max}$, where $f_{\max}=v_{\mathrm{rel}}f_{\mathrm c}/c_0$ is the link-dependent maximum Doppler magnitude, $v_{\mathrm{rel}}=\|\mathbf v_{\mathrm{Tx}}-\mathbf v_{\mathrm{Rx}}\|_2$ is the relative platform speed, $f_{\mathrm c}$ is the carrier frequency, and $c_0$ is the speed of light. The platform velocities are obtained from the time-matched RTK trajectories; because the Rx is fixed in the measurement, $v_{\mathrm{rel}}$ is determined by the UAV velocity. The same Tx velocity is assigned to the RT link when obtaining $\nu_i^{\mathrm{RT}}$. The marginal distributions and Gaussian copula dependence of these parameters are obtained from measurements in Section~\ref{sec:statistical_modeling}.

Given the residual-NLoS power allocation above, the remaining LoS-plus-tail power $P_{\mathrm{L+T}}$ is divided into the LoS and LoS-tail parts as
\begin{equation}
\begin{aligned}
    P_{\mathrm{T}}
    &=
    \eta_{\mathrm{T}}P_{\mathrm{L+T}},\\
    P_{\mathrm{L}}
    &=
    \left(1-\eta_{\mathrm{T}}\right)P_{\mathrm{L+T}} .
\end{aligned}
\label{eq:los_tail_power_split}
\end{equation}
Together with $P_{\mathrm{N}}$, this gives $P_{\mathrm{L}}+P_{\mathrm{T}}+P_{\mathrm{N}}=P_{\mathrm{tot}}$. The LoS component retains the RT delay, Doppler shift, and phase, while $\eta_{\mathrm{T}}$ determines the power assigned to the LoS-tail component.

The LoS-tail component combines the RT-resolved paths selected by $\chi_i^{\mathrm{T,RT}}$ with statistically generated paths. Let $N_{\mathrm{T,RT}}=\sum_{i=1}^{N_{\mathrm{RT}}}\chi_i^{\mathrm{T,RT}}$ and $N_{\mathrm{gen}}=[N_{\mathrm{T}}-N_{\mathrm{T,RT}}]^+$, where $[x]^+=\max(x,0)$. The realized LoS-tail path count is therefore $N_{\mathrm{T}}^{\mathrm{RS}}=N_{\mathrm{T,RT}}+N_{\mathrm{gen}}=\max\{N_{\mathrm{T}},N_{\mathrm{T,RT}}\}$. Thus, $N_{\mathrm T}$ is a sampled augmentation target rather than a strict output count: paths are added only when it exceeds the number already resolved by RT, and existing RT-resolved paths are not removed. The auxiliary parameter $\lambda_{\tau,\mathrm T}>0$ denotes the exponential excess-delay scale used to generate candidate path locations. The implementation sets $\lambda_{\tau,\mathrm T}=\sigma_{\tau,\mathrm T}$ as an initial scale and subsequently optimizes the powers of the combined RT-resolved and generated tail paths to approach the sampled RMS delay and Doppler spreads. For each generated path $r=1,\ldots,N_{\mathrm{gen}}$, the excess delay, Doppler offset, shadowing term, and phase follow
\begin{equation}
\begin{aligned}
    \Delta\tau_r^{\mathrm{gen}}
    &\sim
    \mathrm{Exp}(\lambda_{\tau,\mathrm T})
    \ \big|\ 
    0<\Delta\tau_r^{\mathrm{gen}}\leq\tau_{\mathrm{T}},\\
    \Delta\nu_r^{\mathrm{gen}}
    &\sim
    \mathcal{N}(0,\sigma_{\nu,\mathrm{T}}^2)
    \ \big|\
    \left|\nu_{\mathrm L}+\Delta\nu_r^{\mathrm{gen}}\right|
    \leq f_{\max},\\
    Z_r
    &\sim
    \mathcal{N}(0,\zeta_{\mathrm{T}}^2),
    \qquad
    \phi_r\sim\mathcal{U}(0,2\pi).
\end{aligned}
\label{eq:los_tail_path_generation}
\end{equation}
Here, $\mathrm{Exp}(\lambda_{\tau,\mathrm T})$ has scale parameter $\lambda_{\tau,\mathrm T}$ before truncation by $\tau_{\mathrm T}$, and $Z_r$ and the fixed standard deviation $\zeta_{\mathrm{T}}\geq0$ are expressed in decibels. In the implementation, $\zeta_{\mathrm{T}}$ is set to $3.00~\mathrm{dB}$, following the per-cluster shadowing standard deviation commonly used in the Third Generation Partnership Project (3GPP) Technical Report (TR) 38.901 clustered channel model~\cite{3gpp38901}. The generated locations are referenced to the RT LoS path as $\tau_r^{\mathrm{gen}}=\tau_{\mathrm{L}}+\Delta\tau_r^{\mathrm{gen}}$ and $\nu_r^{\mathrm{gen}}=\nu_{\mathrm{L}}+\Delta\nu_r^{\mathrm{gen}}$. Doppler samples outside the link-dependent interval $[-f_{\max},f_{\max}]$ are rejected and resampled. The generated-path powers decay with excess delay, while $Z_r$ describes random power variation within the tail; conditioned on $\boldsymbol{\Theta}_{\mathrm{T}}$, the generated path variables are sampled independently.

The exponential power decay and random path phases are consistent with conventional cluster-based multipath models~\cite{saleh1987statistical}; this model additionally uses the RT LoS path as the delay and Doppler reference and limits the excess delay to $\tau_{\mathrm{T}}$.

The unnormalized power weight of a generated tail path is modeled as
\begin{equation}
    g_r
    =
    \exp\left(-\frac{\Delta\tau_r^{\mathrm{gen}}}{\lambda_{\tau,\mathrm T}}\right)
    10^{-Z_r/10},
\label{eq:los_tail_generated_weight}
\end{equation}
where $\zeta_{\mathrm{T}}$ controls the lognormal shadowing variation among generated paths. The LoS-tail component is represented by
\begin{equation}
\begin{split}
    h_{\mathrm{T}}(t,\tau)
    =\mathcal{A}_{\mathrm{T}}\Big(&
    \{a_i^{\mathrm{RT}},\tau_i^{\mathrm{RT}},\nu_i^{\mathrm{RT}}\}_{\chi_i^{\mathrm{T,RT}}=1};\\
    &\boldsymbol{\Theta}_{\mathrm{T}},P_{\mathrm{T}},
    \tau_{\mathrm{L}},\nu_{\mathrm{L}},\tau_{\mathrm{T}},
    \lambda_{\tau,\mathrm T},\zeta_{\mathrm{T}}
    \Big).
\end{split}
\label{eq:los_tail_power_operator}
\end{equation}
The operator $\mathcal{A}_{\mathrm{T}}$ preserves available RT-resolved tail locations and phases, generates additional paths when $N_{\mathrm T}>N_{\mathrm{T,RT}}$, and reallocates the combined tail powers to sum to $P_{\mathrm T}$ and approach $(\sigma_{\tau,\mathrm T},\sigma_{\nu,\mathrm T})$. Its realization is specified in Section~\ref{sec:channel_realization}.

\subsection{Channel Realization and Power Optimization}
\label{sec:channel_realization}

\begin{figure}[htbp]
\centering
\resizebox{0.95\columnwidth}{!}{%
\begin{tikzpicture}[
    font=\scriptsize,
    >=Latex,
    node distance=0.30cm and 0.05cm,
    block/.style={draw, rounded corners, align=center, minimum height=0.62cm, text width=5.10cm, fill=gray!6},
    branch/.style={draw, rounded corners, align=center, minimum height=0.72cm, text width=3.15cm},
    los/.style={branch, fill=blue!6},
    nlos/.style={branch, fill=orange!10},
    synthblock/.style={draw, rounded corners, align=center, minimum height=0.76cm, text width=4.90cm, fill=gray!8},
    outblock/.style={draw, rounded corners, align=center, minimum height=0.68cm, text width=4.70cm, fill=green!8},
    decision/.style={draw, diamond, aspect=2.1, align=center, inner sep=0.5pt, text width=2.70cm, fill=gray!8}
]
\node[block] (input) {\textbf{Inputs}\\calibrated RT paths $\{a_i^{\mathrm{RT}},\tau_i^{\mathrm{RT}},\nu_i^{\mathrm{RT}},\ell_i\}$ and $P_{\mathrm{tot}}$\\fitted marginals and Gaussian copula models};
\node[decision, below=of input] (decide) {RT-LoS link?};
\node[los, below left=0.48cm and -0.24cm of decide] (los1) {Sample $\boldsymbol{\Theta}_{\mathrm T}$ and $\boldsymbol{\Theta}_{\mathrm N}$\\$(\eta_{\mathrm T},\sigma_{\tau,\mathrm T},\kappa_{\nu,\mathrm T},N_{\mathrm T})$\\$(\xi_{\mathrm N},\sigma_{\tau,\mathrm N},\kappa_{\nu,\mathrm N})$};
\node[nlos, below right=0.48cm and -0.24cm of decide] (nlos1) {Sample $\boldsymbol{\Theta}_{\mathrm N}^{\mathrm{NLoS}}$\\$(\sigma_{\tau,\mathrm N}^{\mathrm{NLoS}},\kappa_{\nu,\mathrm N}^{\mathrm{NLoS}})$\\set $P_{\mathrm N}=P_{\mathrm{tot}}$};
\node[los, below=of los1] (los2) {Allocate component powers\\$P_{\mathrm N}=\xi_{\mathrm N}P_{\mathrm{tot}}$\\$P_{\mathrm T}=\eta_{\mathrm T}(P_{\mathrm{tot}}-P_{\mathrm N})$\\$P_{\mathrm L}=P_{\mathrm{tot}}-P_{\mathrm N}-P_{\mathrm T}$};
\node[nlos, below=of nlos1] (nlos2) {Set NLoS targets\\$\sigma_{\tau}^{\star}=\sigma_{\tau,\mathrm N}^{\mathrm{NLoS}}$\\$\sigma_{\nu}^{\star}=\kappa_{\nu,\mathrm N}^{\mathrm{NLoS}}f_{\max}$};
\node[los, below=of los2] (los3) {Optimize residual-NLoS powers\\$\sigma_{\tau}^{\star}=\sigma_{\tau,\mathrm N}$\\$\sigma_{\nu}^{\star}=\kappa_{\nu,\mathrm N}f_{\max}$\\$\widehat{\boldsymbol{\psi}}=\arg\min J_{\mathrm N}(\boldsymbol{\psi})$};
\node[nlos, below=of nlos2] (nlos3) {Optimize all RT path powers\\normalize $(x_{\tau,i},x_{\nu,i})$\\$\widehat{\boldsymbol{\psi}}=\arg\min J_{\mathrm N}(\boldsymbol{\psi})$};
\node[los, below=of los3] (los4) {Generate and optimize LoS-tail paths\\$N_{\mathrm{gen}}=[N_{\mathrm T}-N_{\mathrm{T,RT}}]^+$\\sample path support; optimize combined powers};
\node[synthblock, below right=0.48cm and -1.68cm of los4] (synth) {\textbf{Construct channel realization}\\RT-LoS: rescale the LoS path and combine components\\RT-NLoS: retain RT locations and phases and assign optimized powers};
\node[outblock, below=of synth] (out) {\textbf{Output}\\RS-ART channel $h^{\mathrm{RS\text{-}ART}}(t,\tau)$\\and multipath parameters};
\draw[->] (input) -- (decide);
\draw[->] (decide) -- node[above left] {Yes} (los1);
\draw[->] (decide) -- node[above right] {No} (nlos1);
\draw[->] (los1) -- (los2);
\draw[->] (nlos1) -- (nlos2);
\draw[->] (los2) -- (los3);
\draw[->] (nlos2) -- (nlos3);
\draw[->] (los3) -- (los4);
\draw[->] (los4) -- (synth);
\draw[->] (nlos3) -- (synth);
\draw[->] (synth) -- (out);
\end{tikzpicture}}
\caption{RS-ART channel realization flow.}
\label{fig:hybrid_simulation_flowchart}
\end{figure}
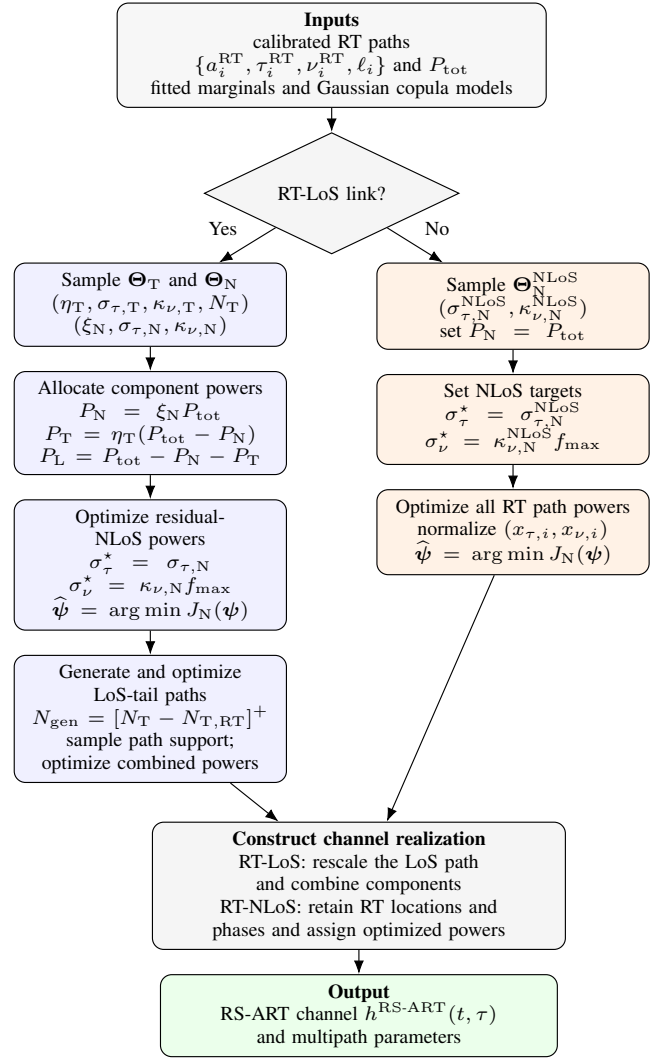

This subsection gives the numerical realization of the NLoS and LoS-tail operators in~\eqref{eq:nlos_power_operator} and~\eqref{eq:los_tail_power_operator}. Fig.~\ref{fig:hybrid_simulation_flowchart} expands the channel-realization stage in Fig.~\ref{fig:rs_art_framework} into parameter sampling, component-power allocation, path generation, and power optimization for RT-LoS and RT-NLoS links.

\emph{Step 1 (Parameter Sampling and Power Allocation):}
For an RT-LoS link, $\boldsymbol{\Theta}_{\mathrm{N}}$ and $\boldsymbol{\Theta}_{\mathrm{T}}$, defined in~\eqref{eq:nlos_parameter_vectors} and~\eqref{eq:los_tail_parameter_vector}, are sampled from their fitted marginals and Gaussian copula models. The NLoS component power is first obtained from~\eqref{eq:nlos_power_allocation}, after which~\eqref{eq:los_tail_power_split} determines $P_{\mathrm{T}}$ and $P_{\mathrm{L}}$. For an RT-NLoS link, only $\boldsymbol{\Theta}_{\mathrm{N}}^{\mathrm{NLoS}}$ is sampled, and~\eqref{eq:nlos_power_allocation} gives $P_{\mathrm{N}}=P_{\mathrm{tot}}$.

\emph{Step 2 (NLoS Power Optimization):}
Let $\bar{\tau}_{\mathrm{N}}$ and $\bar{\nu}_{\mathrm{N}}$ denote the unweighted mean delay and Doppler shift of the retained NLoS paths, and let $s_{\tau,\mathrm{RT}}$ and $s_{\nu,\mathrm{RT}}$ denote their unweighted RMS spreads. The nonnegative late-delay offset is $d_{\tau,i}=\tau_i^{\mathrm{RT}}-\tau_{\mathrm{L}}$ for an RT-LoS link and $d_{\tau,i}=\tau_i^{\mathrm{RT}}-\min_{j\in\mathcal{S}_{\mathrm{N}}}\tau_j^{\mathrm{RT}}$ for an RT-NLoS link. Given the state-dependent targets $(\sigma_{\tau}^{\star},\sigma_{\nu}^{\star})$ supplied to the operator in~\eqref{eq:nlos_power_operator}, the normalized coordinates are
\begin{equation}
\begin{aligned}
    c_{\tau,\mathrm{N}}
    &=\max\{s_{\tau,\mathrm{RT}},\sigma_{\tau}^{\star},\epsilon_{\tau}\},
    &
    x_{\tau,i}
    &=\frac{\tau_i^{\mathrm{RT}}-\bar{\tau}_{\mathrm{N}}}{c_{\tau,\mathrm{N}}},\\
    c_{\nu,\mathrm{N}}
    &=\max\{s_{\nu,\mathrm{RT}},\sigma_{\nu}^{\star},\epsilon_{\nu}\},
    &
    x_{\nu,i}
    &=\frac{\nu_i^{\mathrm{RT}}-\bar{\nu}_{\mathrm{N}}}{c_{\nu,\mathrm{N}}}.
\end{aligned}
\end{equation}
Here, $\epsilon_{\tau}$ and $\epsilon_{\nu}$ are positive numerical floors. Define the normalized RT power within the retained NLoS set as $q_i^{\mathrm{RT}}=P_i^{\mathrm{RT}}/\sum_{j\in\mathcal{S}_{\mathrm{N}}}P_j^{\mathrm{RT}}$. The power-shaping weights and the resulting NLoS powers are
\begin{equation}
\begin{aligned}
    w_i^{\mathrm{N}}(\boldsymbol{\psi})
    &=
    \left(q_i^{\mathrm{RT}}\right)^{\gamma_{\mathrm{P}}}
    \exp\left(
    -\frac{d_{\tau,i}}{\tau_{\mathrm{d}}}
    -\psi_{\tau}x_{\tau,i}^{2}
    -\psi_{\nu}x_{\nu,i}^{2}
    \right),\\
    P_i^{\mathrm{N}}(\boldsymbol{\psi})
    &=
    \frac{w_i^{\mathrm{N}}(\boldsymbol{\psi})}
    {\sum_{j\in\mathcal{S}_{\mathrm{N}}}w_j^{\mathrm{N}}(\boldsymbol{\psi})}
    P_{\mathrm{N}},
    \qquad i\in\mathcal{S}_{\mathrm{N}}.
\end{aligned}
\end{equation}
The exponent $\gamma_{\mathrm{P}}\geq0$ controls the preservation of the RT power ordering, the decay constant $\tau_{\mathrm{d}}>0$ controls late-delay attenuation, and $\boldsymbol{\psi}=[\psi_{\tau},\psi_{\nu}]^{\mathrm T}$ adjusts the two spreads. Let $\sigma_{\tau}^{\mathrm{RS}}(\boldsymbol{\psi})$ and $\sigma_{\nu}^{\mathrm{RS}}(\boldsymbol{\psi})$ be the weighted RMS spreads obtained from $q_i^{\mathrm{N}}=P_i^{\mathrm{N}}/P_{\mathrm{N}}$. The shaping parameter is obtained from
\begin{equation}
\begin{aligned}
    e_{\tau}(\boldsymbol{\psi})
    &=
    \log\frac{\max\{\sigma_{\tau}^{\mathrm{RS}}(\boldsymbol{\psi}),\epsilon_{\tau}\}}
    {\max\{\sigma_{\tau}^{\star},\epsilon_{\tau}\}},\\
    e_{\nu}(\boldsymbol{\psi})
    &=
    \log\frac{\max\{\sigma_{\nu}^{\mathrm{RS}}(\boldsymbol{\psi}),\epsilon_{\nu}\}}
    {\max\{\sigma_{\nu}^{\star},\epsilon_{\nu}\}},\\
    \widehat{\boldsymbol{\psi}}
    &=
    \arg\min_{\boldsymbol{\psi}\in[-\psi_{\max},\psi_{\max}]^2}
    J_{\mathrm{N}}(\boldsymbol{\psi}),\\
    J_{\mathrm{N}}(\boldsymbol{\psi})
    &=
    e_{\tau}^{2}(\boldsymbol{\psi})
    +e_{\nu}^{2}(\boldsymbol{\psi})
    +\rho\|\boldsymbol{\psi}\|_2^2 .
\end{aligned}
\end{equation}
The first two terms quantify the logarithmic differences between the realized and sampled delay and Doppler spreads, while $\rho\geq0$ regularizes the shaping coefficients and $\psi_{\max}>0$ bounds the search region. The implementation uses $\gamma_{\mathrm{P}}=0.65$, $\tau_{\mathrm{d}}=250.00~\mathrm{ns}$, $\epsilon_{\tau}=10^{-12}~\mathrm{s}$, $\epsilon_{\nu}=10^{-9}~\mathrm{Hz}$, $\rho=10^{-3}$, and $\psi_{\max}=10$.

The optimized powers $P_i^{\mathrm{N}}(\widehat{\boldsymbol{\psi}})$ retain the RT locations and phases. If fewer than two paths are available, the calibrated RT powers are normalized to sum to $P_{\mathrm{N}}$ without spread adjustment. If $\mathcal{S}_{\mathrm{N}}$ is empty for an RT-LoS link, $P_{\mathrm{N}}$ is set to zero, the corresponding power is reassigned to $P_{\mathrm{L+T}}$, and the LoS-tail power split is recomputed. An RT-NLoS result without a valid path is excluded because it does not provide an RT channel realization to reshape.

\emph{Step 3 (LoS-Tail Generation and Power Optimization):}
This step is applied only to RT-LoS links. The LoS coefficient is rescaled to $\sqrt{P_{\mathrm{L}}}e^{j\angle a_{i_{\mathrm{L}}}^{\mathrm{RT}}}$ without changing its delay or Doppler shift. The RT-resolved tail retains its path locations and phases, and $N_{\mathrm{gen}}$ additional paths are sampled according to~\eqref{eq:los_tail_path_generation}. Let $\mathcal S_{\mathrm{T,RT}}=\{i:\chi_i^{\mathrm{T,RT}}=1\}$ and $\mathcal S_{\mathrm{T,gen}}=\{r:1\leq r\leq N_{\mathrm{gen}}\}$ denote the two disjoint path sets. They are combined as $\mathcal S_{\mathrm T}^{+}=\mathcal S_{\mathrm{T,RT}}\cup\mathcal S_{\mathrm{T,gen}}$.

Let $N_{\mathrm T}^{+}=|\mathcal S_{\mathrm T}^{+}|$ and let $b_j$ denote the initial power weight of path $j\in\mathcal S_{\mathrm T}^{+}$, obtained from its RT power or the generated weight $g_r$ in~\eqref{eq:los_tail_generated_weight}. For $y\in\{\tau,\nu\}$, define $y_j=\tau_j$ or $\nu_j$, $\sigma_{y,\mathrm T}=\sigma_{\tau,\mathrm T}$ or $\sigma_{\nu,\mathrm T}$, and $\epsilon_y=\epsilon_\tau$ or $\epsilon_\nu$. The combined path coordinates are normalized as
\begin{equation}
\begin{aligned}
    \overline y_{\mathrm T}
    &=
    \frac{1}{N_{\mathrm T}^{+}}
    \sum_{j\in\mathcal S_{\mathrm T}^{+}}y_j,\\
    s_{y,\mathrm T}^{(0)}
    &=
    \left[
    \frac{1}{N_{\mathrm T}^{+}}
    \sum_{j\in\mathcal S_{\mathrm T}^{+}}
    \left(y_j-\overline y_{\mathrm T}\right)^2
    \right]^{1/2},\\
    c_{y,\mathrm T}
    &=
    \max\left\{
    s_{y,\mathrm T}^{(0)},\sigma_{y,\mathrm T},\epsilon_y
    \right\},\\
    x_{y,j}^{\mathrm T}
    &=
    \left(
    \frac{y_j-\overline y_{\mathrm T}}{c_{y,\mathrm T}}
    \right)^2 .
\end{aligned}
\end{equation}
With $\boldsymbol{\theta}=[\theta_\tau,\theta_\nu]^{\mathrm T}$, the combined tail powers are
\begin{equation}
\begin{aligned}
    w_j^{\mathrm T}(\boldsymbol{\theta})
    &=
    b_j\exp\left(
    \theta_{\tau}x_{\tau,j}^{\mathrm T}
    +\theta_{\nu}x_{\nu,j}^{\mathrm T}
    \right),\\
    P_j^{\mathrm T}(\boldsymbol{\theta})
    &=
    \frac{w_j^{\mathrm T}(\boldsymbol{\theta})}
    {\sum_{m\in\mathcal S_{\mathrm T}^{+}}w_m^{\mathrm T}(\boldsymbol{\theta})}
    P_{\mathrm T},
    \qquad
    j\in\mathcal S_{\mathrm T}^{+}.
\end{aligned}
\end{equation}
Let $\sigma_{y,\mathrm T}^{\mathrm{RS}}(\boldsymbol{\theta})$ denote the power-weighted RMS spread of $\{y_j\}_{j\in\mathcal S_{\mathrm T}^{+}}$ under the normalized powers $P_j^{\mathrm T}(\boldsymbol{\theta})/P_{\mathrm T}$. The spread errors and objective are
\begin{equation}
\begin{aligned}
    e_{y,\mathrm T}(\boldsymbol{\theta})
    &=
    \log
    \frac{
    \max\{\sigma_{y,\mathrm T}^{\mathrm{RS}}(\boldsymbol{\theta}),\epsilon_y\}
    }{
    \max\{\sigma_{y,\mathrm T},\epsilon_y\}
    },
    \qquad y\in\{\tau,\nu\},\\
    \widehat{\boldsymbol{\theta}}
    &=
    \arg\min_{\boldsymbol{\theta}\in[-\theta_{\max},\theta_{\max}]^2}
    J_{\mathrm T}(\boldsymbol{\theta}),\\
    J_{\mathrm T}(\boldsymbol{\theta})
    &=
    e_{\tau,\mathrm T}^{2}(\boldsymbol{\theta})
    +
    e_{\nu,\mathrm T}^{2}(\boldsymbol{\theta})
    +
    \rho_{\mathrm T}\|\boldsymbol{\theta}\|_2^2 .
\end{aligned}
\end{equation}
The implementation uses $\rho_{\mathrm T}=10^{-3}$ and $\theta_{\max}=12$. The optimized tail weights are applied to the combined LoS-tail support. If fewer than two tail paths are available, their powers are only normalized to $P_{\mathrm T}$.

\emph{Step 4 (Channel Synthesis):}
The optimized components are combined according to~\eqref{eq:rs_art_channel}. The resulting channel has total power $P_{\mathrm{tot}}$, with the RT path locations unchanged and the generated LoS-tail paths placed at their sampled delay and Doppler coordinates.

\begin{figure*}[!t]
    \centering
    \subfloat[]{\includegraphics[width=0.65\columnwidth]{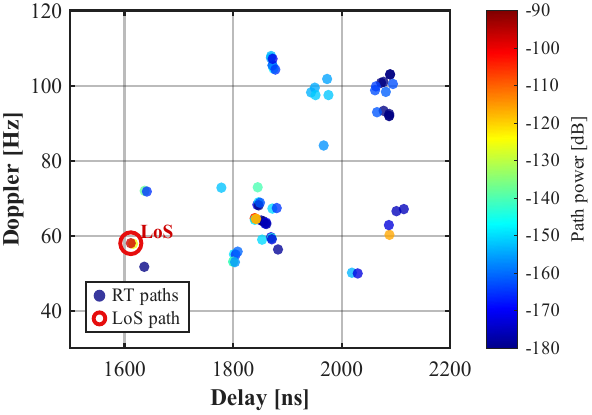}}
    \hfill
    \subfloat[]{\includegraphics[width=0.65\columnwidth]{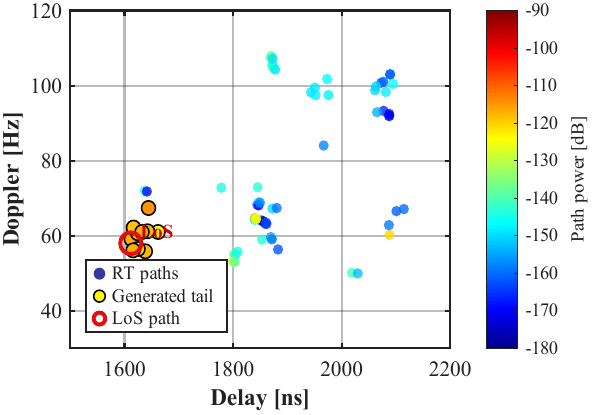}}
    \hfill
    \subfloat[]{\includegraphics[width=0.65\columnwidth]{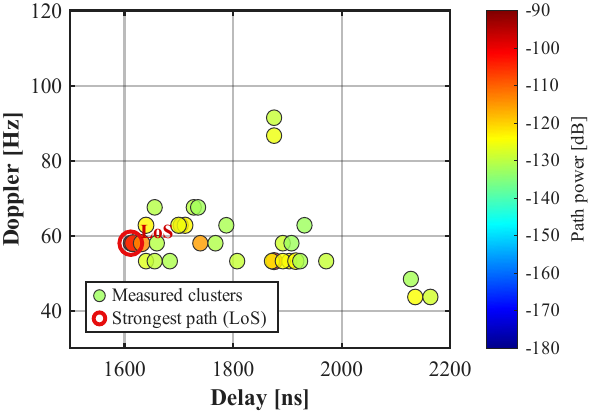}}
    \caption{Path-power comparison over delay and Doppler: (a) calibrated RT; (b) RS-ART; and (c) measurement.}
    \label{fig:hybrid_model_frame}
\end{figure*}

Fig.~\ref{fig:hybrid_model_frame} shows one channel realization. Compared with the calibrated RT result in Fig.~\ref{fig:hybrid_model_frame}(a), the measurement in Fig.~\ref{fig:hybrid_model_frame}(c) contains more paths and multipath power in the LoS-tail delay region. In Fig.~\ref{fig:hybrid_model_frame}(b), RS-ART retains the RT-resolved paths, supplies additional short-delay paths when required by the sampled $N_{\mathrm T}$, and adjusts the component and path powers using the sampled power-ratio and spread parameters. The measurement is included as a qualitative reference because its paths are not in one-to-one correspondence with the RT realization.

The evaluation uses the same 1079 position-matched links as the RT calibration. For the CDF comparisons in Figs.~\ref{fig:validation_los_tail_cdf}--\ref{fig:validation_nlos_cdf}, one independent RS-ART realization is generated for each position-matched RT link, and the resulting statistics over all applicable positions form the empirical RS-ART CDFs. The 20-realization averaging is used only for the reported delay spread and Doppler spread RMSEs. Specifically, $K=20$ independent RS-ART realizations are generated from the RT result associated with each measured position. Let $g_m^{(k)}$ denote a channel statistic from realization $k$ at position $m$, and let $g_m^{\mathrm{meas}}$ denote its measured value. The realization-averaged statistic and its location-matched RMSE over $M$ positions are
\begin{equation}
\begin{aligned}
\overline g_m&=\frac{1}{K}\sum_{k=1}^{K}g_m^{(k)},\\
\operatorname{RMSE}(g)&=\sqrt{\frac{1}{M}\sum_{m=1}^{M}
\left(\overline g_m-g_m^{\mathrm{meas}}\right)^2}.
\end{aligned}
\end{equation}
The CDFs and location-matched RMSEs compare different quantities. The CDFs compare the distributions obtained from one realization over the measurement route. The location-matched RMSE instead compares the statistic averaged over 20 realizations with the measured value at the same Tx--Rx position. Averaging reduces the effect of an individual random draw, but the sampled parameters are still not matched to the instantaneous measured multipath.

\subsection{Validation for RT-LoS Links}

\begin{figure}
		\centering
		\subfloat[$\eta_{\mathrm{T}}$]{\includegraphics[width=0.49\columnwidth]{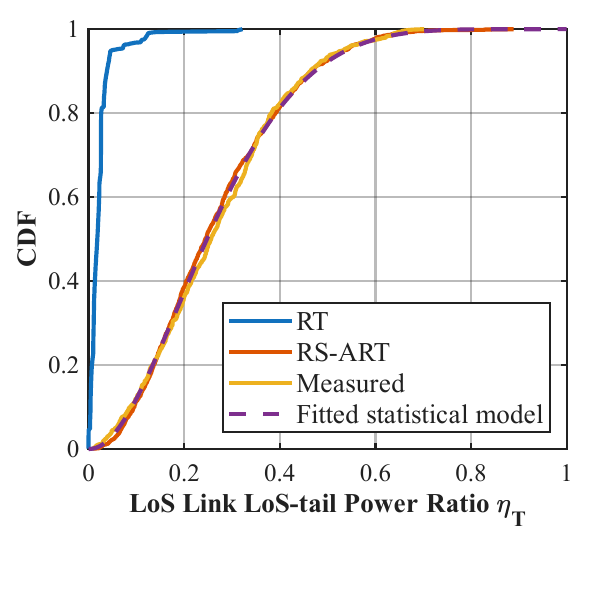}}
		\hfill
		\subfloat[$\sigma_{\tau,\mathrm{T}}$]{\includegraphics[width=0.49\columnwidth]{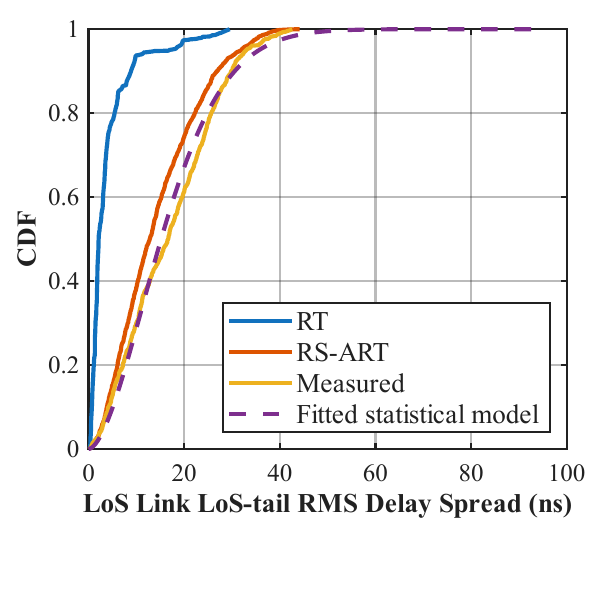}}\\[0.6em]
		\subfloat[$\kappa_{\nu,\mathrm{T}}$]{\includegraphics[width=0.49\columnwidth]{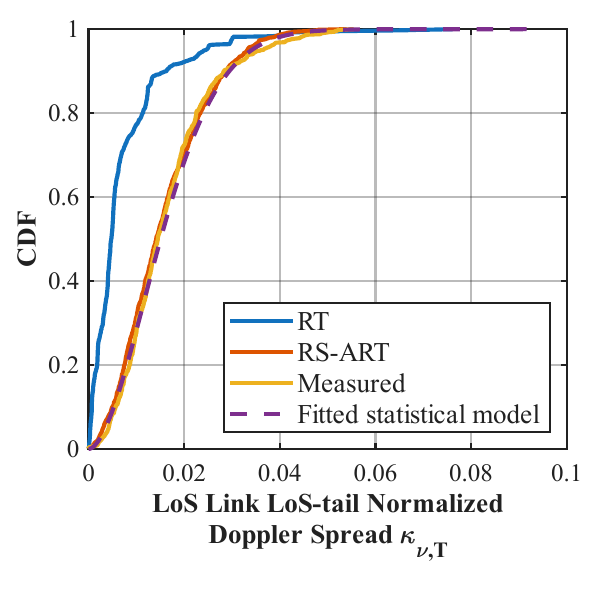}}
		\hfill
		\subfloat[$N_{\mathrm{T}}$]{\includegraphics[width=0.49\columnwidth]{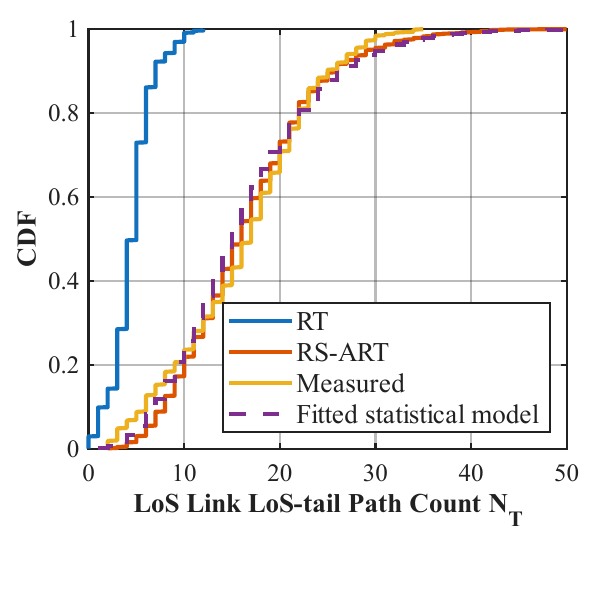}
			\label{fig:validation_los_tail_count_cdf}}
		\caption{CDFs of the LoS-tail statistics for RT-LoS links.}
		\label{fig:validation_los_tail_cdf}
\end{figure}

For RT-LoS links, Fig.~\ref{fig:validation_los_tail_cdf} compares the CDFs of $\eta_{\mathrm T}$, $\sigma_{\tau,\mathrm T}$, $\kappa_{\nu,\mathrm T}$, and $N_{\mathrm T}$. At a CDF of $0.50$, the calibrated RT medians are approximately $0.01$, $2$~ns, $0.005$, and 4 paths, respectively, whereas the measured medians are approximately $0.26$, $17$~ns, $0.014$, and 17 paths. The corresponding RS-ART medians are approximately $0.19$, $13$~ns, $0.014$, and 13 paths. The remaining differences at the median are therefore approximately $0.07$, $4$~ns, $0$, and 4 paths. The generated paths increase the short-delay path count, and the power adjustment assigns more power to this region to compensate for local and time-varying objects omitted from the 3D RT scene.

For all four LoS-tail statistics, the RS-ART CDF is closer to the measured CDF than the calibrated RT CDF over the middle probability range. At a CDF of $0.50$, the RS-ART and measured normalized Doppler spreads are both $0.014$ to the reported precision. The median differences remain $0.07$ for the tail power ratio, $4$~ns for the RMS delay spread, and 4 paths for the path count. RS-ART retains the delays of the original RT paths, so these delays continue to affect the realized delay spread. The power ratio and path count are also sampled from fitted distributions rather than assigned the measured value of each frame. These two factors account for the remaining differences in the medians and distribution tails.

Fig.~\ref{fig:validation_los_residual_nlos_cdf} compares the residual-NLoS statistics for RT-LoS links. The median $\xi_{\mathrm N}$ decreases from approximately $0.26$ for calibrated RT to $0.06$ for RS-ART; the measured median is also approximately $0.06$. For $\sigma_{\tau,\mathrm N}$, the RS-ART CDF lies between the measured and fitted curves over approximately $50$--$200$~ns. For $\kappa_{\nu,\mathrm N}$, the median increases from approximately $0.02$ for calibrated RT to $0.05$ for RS-ART, compared with approximately $0.06$ for the measurement. The RS-ART and measured CDFs remain separated above $200$~ns and $\kappa_{\nu,\mathrm N}>0.20$. The power optimization changes path powers but not the RT delay and Doppler locations.

{\setlength{\intextsep}{6pt}
\begin{figure}[!t]
    \centering
    \subfloat[$\xi_{\mathrm{N}}$]{\includegraphics[width=0.48\columnwidth]{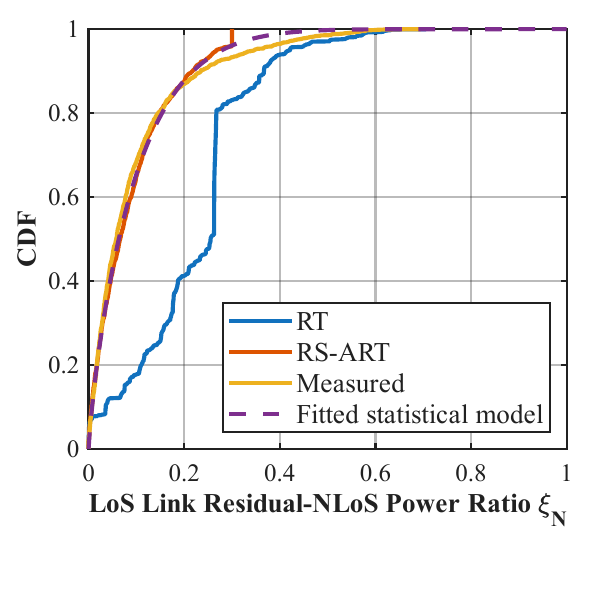}}
    \hfill
    \subfloat[$\sigma_{\tau,\mathrm{N}}$]{\includegraphics[width=0.48\columnwidth]{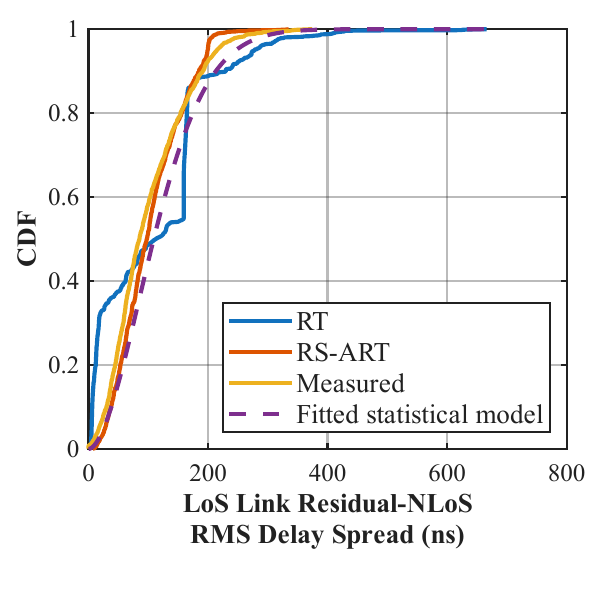}}\\[0.6em]
    \subfloat[$\kappa_{\nu,\mathrm{N}}$]{\includegraphics[width=0.48\columnwidth]{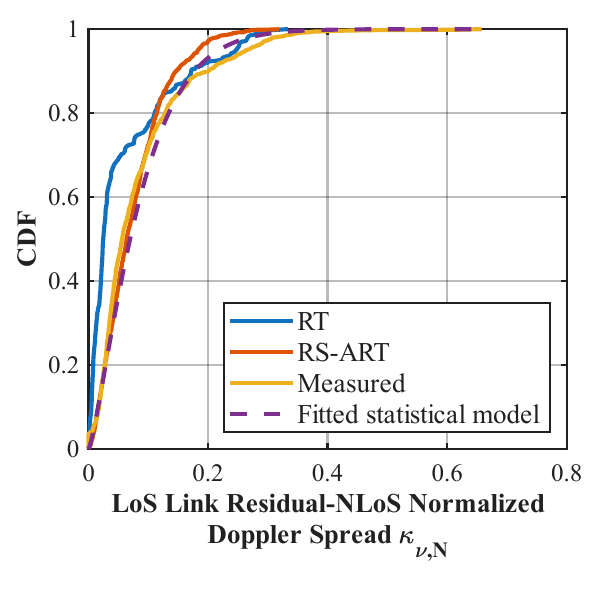}}
    \caption{CDFs of the residual-NLoS statistics for RT-LoS links.}
    \label{fig:validation_los_residual_nlos_cdf}
\end{figure}
}

After the LoS-tail and residual-NLoS components are evaluated separately, Fig.~\ref{fig:validation_los_overall_cdf} compares the channel-level RMS delay spread and normalized Doppler spread of RT-LoS links. In Fig.~\ref{fig:validation_los_overall_cdf}(b), the calibrated RT CDF is closer to the measured CDF over part of the probability range. The CDF pools all positions and does not compare the models at the same Tx--Rx position, so this visual result does not imply a smaller position-matched error. The RS-ART CDF also uses only one random realization at each position. By contrast, the location-matched RMSE compares the measurement with the mean of 20 RS-ART realizations generated at the same position. In this comparison, RS-ART reduces the RMS delay spread RMSE from $87.64$~ns to $42.76$~ns ($51.21\%$) and the normalized Doppler spread RMSE from $0.0879$ to $0.0682$ ($22.36\%$). Thus, calibrated RT is closer in part of the aggregated Doppler CDF, while RS-ART has the lower position-matched RMSE.

{\setlength{\intextsep}{6pt}
\begin{figure}[!t]
    \centering
    \subfloat[RMS delay spread]{\includegraphics[width=0.48\columnwidth]{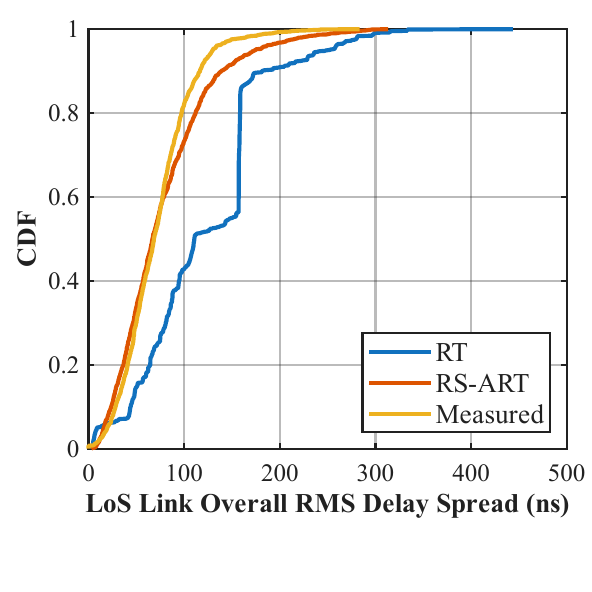}}
    \hfill
    \subfloat[Normalized Doppler spread]{\includegraphics[width=0.48\columnwidth]{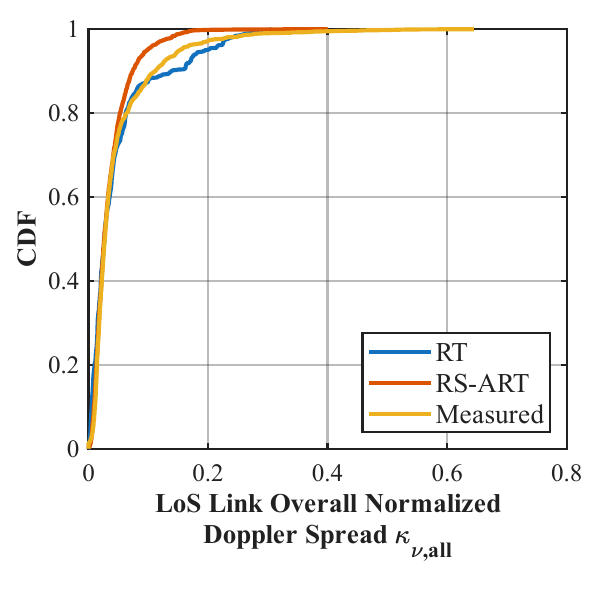}}
    \caption{CDFs of the channel-level statistics for RT-LoS links.}
    \label{fig:validation_los_overall_cdf}
\end{figure}
}

\subsection{Validation for RT-NLoS Links}

For RT-NLoS links, the NLoS component represents the full channel. Fig.~\ref{fig:validation_nlos_cdf} compares its RMS delay spread and normalized Doppler spread. In Fig.~\ref{fig:validation_nlos_cdf}(a), the RS-ART curve lies slightly to the right of the measured curve over most of the $50$--$200$~ns interval, whereas calibrated RT contains more values below $100$~ns and above $200$~ns. In Fig.~\ref{fig:validation_nlos_cdf}(b), the RS-ART curve lies to the right of the measured curve for CDF values from approximately $0.20$ to $0.90$, while calibrated RT retains values above $0.10$. The location-matched RMS delay spread RMSE decreases from $126.31$~ns to $55.40$~ns ($56.14\%$), and the normalized Doppler spread RMSE decreases from $0.0657$ to $0.0267$ ($59.38\%$). All retained path powers are optimized together, while their RT delay and Doppler locations remain unchanged.

{\setlength{\intextsep}{6pt}
\begin{figure}[!t]
    \centering
    \subfloat[RMS delay spread]{\includegraphics[width=0.48\columnwidth]{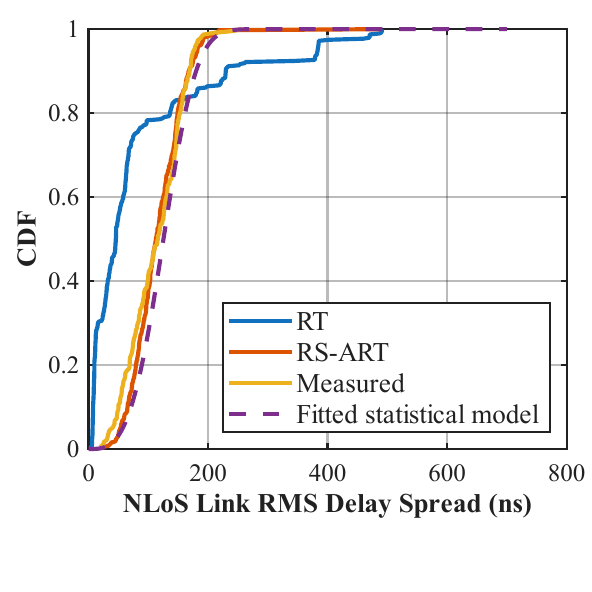}}
    \hfill
    \subfloat[Normalized Doppler spread]{\includegraphics[width=0.48\columnwidth]{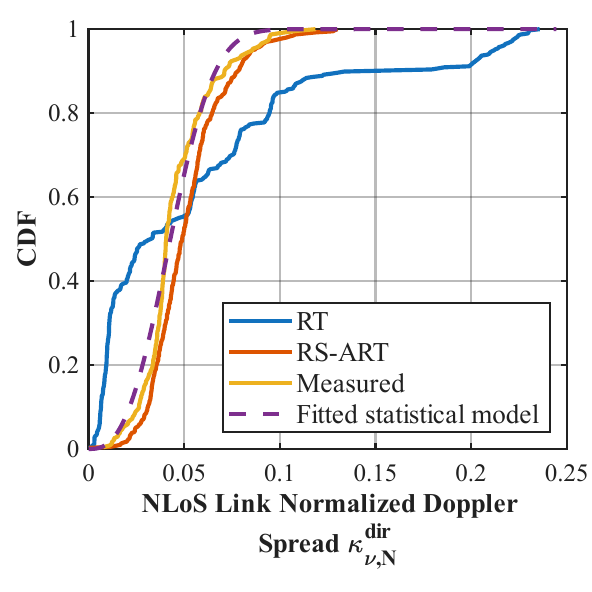}}
    \caption{CDFs of the NLoS statistics for RT-NLoS links.}
    \label{fig:validation_nlos_cdf}
\end{figure}
}

Table~\ref{tab:overall_location_rmse} consolidates the RMSE results reported above. It lists the input RMSE, output RMSE, and percentage reduction for path loss calibration and RS-ART spread validation, with the results separated into all matched, RT-LoS, and RT-NLoS links.
{
\setlength{\intextsep}{6pt}
\begin{table}[H]
\centering
\caption{RMSE summary for RT calibration and RS-ART augmentation.}
\label{tab:overall_location_rmse}
\resizebox{\columnwidth}{!}{%
\begin{tabular}{llccc}
\toprule
Link set & Metric & Input RMSE & Output RMSE & Reduction \\
\midrule
\multicolumn{5}{l}{\textit{RT calibration: original RT $\rightarrow$ calibrated RT}}\\
All matched & Path loss [dB] & 5.45 & 4.35 & 20.18\% \\
RT-LoS & Path loss [dB] & 4.10 & 4.05 & 1.22\% \\
RT-NLoS & Path loss [dB] & 5.91 & 3.13 & 47.04\% \\
\midrule
\multicolumn{5}{l}{\textit{RS-ART augmentation: calibrated RT $\rightarrow$ RS-ART}}\\
All matched & RMS delay spread [ns] & 97.85 & 45.96 & 53.03\% \\
All matched & Normalized Doppler spread & 0.0833 & 0.0612 & 26.48\% \\
\midrule
RT-LoS & RMS delay spread [ns] & 87.64 & 42.76 & 51.21\% \\
RT-LoS & Normalized Doppler spread & 0.0879 & 0.0682 & 22.36\% \\
\midrule
RT-NLoS & RMS delay spread [ns] & 126.31 & 55.40 & 56.14\% \\
RT-NLoS & Normalized Doppler spread & 0.0657 & 0.0267 & 59.38\% \\
\bottomrule
\end{tabular}%
}
\end{table}
}

Overall, electromagnetic calibration reduces the path loss RMSE from $5.45$ to $4.35$~dB. Across the position-matched links, RS-ART reduces the RMS delay spread RMSE from $97.85$ to $45.96$~ns and the normalized Doppler spread RMSE from $0.0833$ to $0.0612$, corresponding to reductions of $53.03\%$ and $26.48\%$, respectively.

\section{Conclusion}

This paper presents the RS-ART framework to address two limitations of site-specific RT channel modeling: the difficulty of obtaining 3D scene inputs over wide areas and the incomplete representation of local, time-varying multipath in 3D RT scenes. Satellite remote-sensing imagery provides the input for 3D scene construction, while measurement-derived channel statistics supplement the RT results. This combination reduces the dependence on pre-existing 3D maps while retaining the path delays and Doppler locations calculated for each RT link.
Analysis of the 4.60~GHz UAV measurements using the proposed multipath extraction method shows that the measured channels contain more short-delay paths near the LoS component than the RT results. The fitted dependence models indicate a positive relationship between the LoS-tail delay and normalized Doppler spreads, whereas the residual-NLoS parameters exhibit weaker dependence. Based on these observations, RS-ART supplements RT-LoS links with statistically generated LoS-tail paths and adjusts the path powers for both RT-LoS and RT-NLoS links, reducing the differences in the resulting channel statistics.

The validation results show that the framework reduces the path loss RMSE from $5.45$ to $4.35$~dB and decreases the RMS delay spread and normalized Doppler spread RMSEs by $53.03\%$ and $26.48\%$, respectively. These results show that RS-ART reduces the differences between the site-specific RT results and the measured channel statistics while using remote-sensing imagery as the scene input. The proposed framework provides a site-specific channel modeling approach for 6G space--air--ground digital-twin studies.

\bibliographystyle{IEEEtran}
\bibliography{mybib}

\end{document}